\documentclass[a4paper,11pt]{article}

\pdfoutput=1 
\usepackage{jcappub} 
\usepackage{lineno}
\usepackage{graphicx}
\usepackage{subcaption}
\usepackage{mathrsfs}
\usepackage{bm} 
\usepackage{tablefootnote}
\usepackage{aas_macros}
\usepackage{subcaption}  
\usepackage{hyperref}
\usepackage{amssymb,amsmath}
\usepackage[dvipsnames]{xcolor}
\usepackage[normalem]{ulem}
\usepackage{makecell}

\newcommand{\eqn}[1]{equation~(\ref{#1})}

\newcommand{\secn}[1]{Section~\ref{#1}}
\newcommand{\secns}[2]{Section~\ref{#1} and \ref{#2}}
\newcommand{\fig}[1]{Figure~\ref{#1}}

\newcommand{\tab}[1]{Table~\ref{#1}}
\newcommand{\app}[1]{Appendix~\ref{#1}}
\newcommand{\angstrom}{\text{\normalfont\AA}}
\newcommand{\Msun}{\rm{M_\odot}}

\newcommand{\MUV}{M_{\rm UV}}

\newcommand{\kappaUV} {\mathcal{K}_{{\rm UV}}}
\newcommand{\kappaUVfid} {\mathcal{K}_{{\rm UV,fid}}}

\newcommand{\hcMpc}{h^{-1}~\mathrm{cMpc}}
\newcommand{\cubedhcMpc}{h^{-3}~\mathrm{cMpc}^3}

\defcitealias{Chakraborty2024}{CC24}
\defcitealias{Chakraborty2025}{CC25}
\newcommand{\OIIIang}{\ifmmode \text{[O{\sc iii}]}~5008\angstrom \else[O{\sc iii}] 5008\angstrom \fi}
\newcommand{\OIII}{\ifmmode \text{[O{\sc iii}]} \else[O{\sc iii}]\fi}
\newcommand{\CfacOIII} {\mathcal{C}_{\OIII}}

\arxivnumber{2602.15803} 
\title{Nearest Neighbour-Based Statistics for 21\,cm-Galaxy Cross-Correlations in the Epoch of Reionization}

\author[a,1]{Anirban Chakraborty, \note{Corresponding author.}}
\author[b,c]{Kwanit Gangopadhyay,}
\author[b]{Arka Banerjee,}
\author[a]{and Tirthankar Roy Choudhury }

\affiliation[a]{National Centre for Radio Astrophysics, Tata Institute of Fundamental Research, \\Pune University Campus, Ganeshkhind, Pune 411007, India.}
\affiliation[b]{Department of Physics, Indian Institute of Science Education and Research,
Homi Bhabha Road, Pashan, Pune 411008, India.}
\affiliation[c]{Van Swinderen Institute for Particle Physics and Gravity, University of Groningen, Nijenborgh 3, NL-9747 AG Groningen, The Netherlands.}

\emailAdd{anirban@ncra.tifr.res.in, anirban.chakraborty096@gmail.com}

\abstract{21\,cm radiation from neutral hydrogen serves as a direct probe of the Epoch of Reionization. However, both its detection and physical interpretation are severely hindered by contamination from astrophysical foreground emission and instrumental systematics that are several orders of magnitude brighter than the signal of interest. A promising way to tackle these challenges is to cross-correlate the 21\,cm signal with other independent tracers of large-scale structure, most notably high-redshift galaxies. Besides validating putative 21\,cm detections, such joint analyses are expected to provide independent insights into the properties of ionizing sources and the evolving morphology of ionized regions during reionization.
The 21\,cm signal, however, is intrinsically highly non-Gaussian, limiting the effectiveness of conventional two-point cross-correlation statistics, which capture information only up to the second order. In this work, we therefore investigate the utility of $k$-nearest-neighbour cumulative distribution functions ($k$NN CDF), which encode information from the joint clustering at all orders, as an alternative framework for probing 21\,cm--galaxy cross-correlations. Using self-consistently simulated mock 21\,cm fields and a catalog of line-emitting galaxies at $z$ = 7 that satisfy all available observations, we conduct a \textbf{\emph{proof-of-concept}} study comparing the $k$NN CDF formalism and the two-point cross-correlation approach.
We find that the $k$NN CDF statistics outperform the two-point statistics in detecting 21\,cm--galaxy cross-correlations, even in the presence of instrumental noise and aggressive foreground filtering. Moreover, at a fixed global ionized fraction, it is even able to differentiate between reionization models that remain indistinguishable using two-point statistics. These results demonstrate the power and relatively unexplored potential of exploiting higher order statistics for extracting maximal information from 21\,cm--galaxy synergies.
}


\begin{document}
\maketitle
\flushbottom

\section{Introduction}
\label{sec:intro}

Cosmic reionization marks a major phase transition in the evolutionary history of the Universe, during which the predominantly neutral intergalactic medium (IGM) was ionized by the first generation of luminous sources. Our current understanding of the Epoch of Reionization (EoR) is informed by a diverse set of observations, including primary and secondary anisotropies of the cosmic microwave background (CMB) \cite{Planck2018, Reichardt2021, Choudhury2021, Nikolic2023, Cain2025}, imprints of Lyman-$\alpha$ absorption in the spectra of high-redshift galaxies and quasars \cite{Fan2006, Gallerani2006, Becker2015, McGreer2015, Kulkarni2019_IGM, Bosman2018, Bosman2022, Zhu2021, Mason2025, Davies2018, Greig2022, Jin2023, Umeda2023, FredDavies2026}, and measurements of the abundance, clustering, and ionizing properties of early galaxies \cite{Ouchi2010, Sobacchi2015, Choudhury2015, Konno2018, Ouchi2018, Itoh2018, Goto2021, Morales2021, Umeda2025_LAE, Simmonds2024_JADES_xiION, Pahl2025_xiION, Begley2025_xiION}. Despite their success, these probes only yield indirect or spatially-averaged constraints on the ionization state of the IGM. In contrast, the redshifted 21\,cm radiation arising from the hyperfine transition of neutral hydrogen is expected to offer a uniquely powerful window into the EoR \cite{Pritchard2012} and thus detecting this signal is a key objective of many existing as well as upcoming radio interferometers, such as the Low-Frequency Array (LOFAR) \cite{vanHaarlem2013_LOFARinst}, the Murchison
Widefield Array (MWA) \cite{Tingay2013_MWAinst}, the Hydrogen Epoch of Reionization Array (HERA) \cite{DeBoer2017_HERA}, the upgraded Giant Metrewave Radio Telescope (uGMRT) \cite{Gupta2017_GMRTinst}, and the SKA Observatory (SKAO) \cite{Koopmans2015_SKA}. As a direct tracer of the neutral hydrogen density in the IGM, spatial fluctuations in the 21\,cm signal will provide an unprecedented three-dimensional, tomographic view of reionization across cosmic time, enabling detailed investigations of its evolution, morphology, and underlying astrophysics that are inaccessible to other existing observational probes \cite{Mellema2015_tomography}.

In practice, detecting the cosmological 21\,cm signal is an extremely challenging task. The signal itself is intrinsically faint and lies buried deep beneath astrophysical foreground emission (e.g., diffuse Galactic emission and emission from extragalactic point sources) that is several orders of magnitude brighter. This is further compounded by the fact that 21\,cm observations are usually also affected by a range of instrumental effects and observational systematics. An important example is the chromatic response of radio interferometers: although the astrophysical foregrounds vary smoothly with observing frequency, instrumental chromaticity causes the measured emission from fixed directions on the sky to appear spectrally structured. This instrumental mixing of angular (spatial) and frequency information typically spreads the otherwise smooth foregrounds into a characteristic wedge-shaped region, commonly referred to as the ``foreground wedge'', in Fourier space. \cite{Pober2014, Pober2013,Liu2014}. If not properly mitigated, contamination from foregrounds and systematics can bias interpretations of the 21\,cm signal.

In this context, cross-correlation analyses offer a promising complementary pathway for isolating the cosmological 21\,cm signal. By cross-correlating 21\,cm observations with independent tracers of large-scale structure at the same redshift, or with complementary observables at other wavelengths, one can preferentially extract the cosmological information common to both datasets, while suppressing contamination from uncorrelated foregrounds and instrumental systematics. As a result, a large body of work has explored the potential of cross-correlations between the 21\,cm signal from the EoR and other cosmological observables, including high-redshift galaxy surveys \cite{FurlanettoLidz2007, Wyithe2007, Lidz2009, Wiersma2013, Vrbanec2016, Sobacchi2016, Hutter2017, Kubota2018, Vrbanec2020, Weinberger2020, LaPlante2023, Moriwaki2024, GagnonHartman2025, Hutter2025}, line-intensity mapping experiments targeting aggregate emission from unresolved high-redshift galaxies \cite{Lidz2011, Gong2012, Silva2013, Heneka2017, Dumitru2019, Murmu2021, Padmanabhan2022, Fronenberg2024, Roy2025}, and cosmic radiation backgrounds such as the microwave \cite{Salvaterra2005, Alvarez2006, Meerburg2013, Ma2018, Beane2019, Roy2020, Zhou2025}, near-infrared \cite{Fernandez2014, Mao2014}, and X-ray \cite{Liang2016, Ma2018_Xray} backgrounds. 
Among these datasets, in the literature, particular attention has been devoted to exploiting the spatial cross-correlations between the fluctuating 21\,cm signal and the distribution of high-redshift galaxies, not only for enhancing the detectability and confirming the cosmological origin of the 21\,cm signal, but also for cleanly extracting independent information about the EoR itself. These studies have revealed that galaxy–21\,cm cross-correlations are sensitive probes of the timing, topology, and sources of reionization \cite{Lidz2009, Wiersma2013, Park2014, Sobacchi2016, Vrbanec2016, Dumitru2019, Kannan2022, Moriwaki2024, Hutter2023, Hutter2025, Pietschke2026}. Importantly, the feasibility of this strategy has already been demonstrated at lower redshifts ($z \lesssim 1$), where statistically significant detections of the post-reionization 21\,cm signal have been achieved through cross-correlation with galaxy surveys \cite{Masui2013, Anderson2018, Wolz2022, Cunnington2023, Amiri2023}.

However, most existing studies of 21\,cm–galaxy synergies have relied on two-point statistics, such as the cross-power spectrum or the real-space cross-correlation function. Given the intrinsically non-Gaussian nature of the EoR 21\,cm signal, these statistical measures provide an incomplete description, as they are insensitive to the higher-order correlations associated with the patchy morphology of reionization. Recently, the $k$-nearest neighbour ($k$NN) cumulative distribution function (CDF) has been proposed as an alternative summary statistic for auto-clustering and cross-clustering analyses of cosmological datasets \cite{Banerjee2021_kNNIntro, Banerjee2021, Banerjee2023_tracerfieldCross}. The $k$NN CDFs are sensitive to all the connected $N$-point correlation functions in the data but can be evaluated with a computational cost similar to that of computing the two-point function~\cite{Banerjee2021_kNNIntro}. The effectiveness of the $k$NN-CDF framework relative to traditional two-point statistics has been demonstrated at low redshifts through its applications to both simulated datasets \cite{Banerjee2021_kNNIntro, Banerjee2021, Banerjee_kNNmodeling}, including scenarios that incorporate observational noise \cite{Banerjee2023_tracerfieldCross, Chand2025}, as well as on observational data \cite{Wang2022, Anbajagane2023, Gupta2024, Zhou2025_KNN}. More recently, Gangopadhyay et al. (2025) \cite{Gangopadhyay2025} showed that the $k$NN CDFs and their derivatives have direct geometric interpretations in terms of volume of spheres (and their intersections) around discrete tracers (e.g., galaxies), further motivating the use of these summary statistics for probing the topology of ionized bubbles during the EoR.

Motivated by these results, we present a \emph{proof-of-concept} study exploring the potential of the $k$NN-CDF framework as a higher-order summary statistic for 21\,cm–galaxy cross-correlation studies during the EoR. Using simulated 21\,cm and high-redshift galaxy datasets, we compare the performance of the $k$NN CDF formalism with conventional two-point statistics in the presence of instrumental and observational systematics. We further assess the sensitivity of the $k$NN CDF cross-correlation measurements to variations in the ionization topology under realistic observational conditions, and examine its ability to discriminate between different reionization models.

This paper is organised as follows: in \secn{sec:theory_model}, we describe the simulations and theoretical model used in preparing mock datasets of high-redshift galaxies and the fluctuating 21 cm signal from the intergalactic medium, \secn{sec:crosscorr_frameworks} introduces the cross-correlation frameworks considered in this work, including the two-point cross-correlation function and the nearest-neighbour cumulative distribution function, while \secn{sec:detection_prospects} presents our analysis methodology and discusses the prospects for detecting the cross-correlation signals using these frameworks. \secn{sec:diff_reion_models} examines how cross-correlation statistics can be used to distinguish between different reionization scenarios. Finally, we summarise our main findings in \secn{sec:conclusion}.


\section{Simulations and Theoretical Model}
\label{sec:theory_model}

In this section, we describe the simulations and theoretical framework used to construct mock datasets for the cross-correlation analysis. We begin with an overview of the N-body simulations, followed by a description of our semi-analytical model of high-redshift galaxies and the semi-numerical approach employed for modeling cosmic reionization.

\subsection{$N$-body Simulations of Cosmological Structure Formation}
\label{subsec:gadget2_sims}
We ran a dark-matter only $N$-body simulation using the publically available {\tt GADGET-2} \footnote{\href{https://wwwmpa.mpa-garching.mpg.de/gadget/}{\texttt{https://wwwmpa.mpa-garching.mpg.de/gadget/}}} code \cite{gadget2}, assuming a flat $\Lambda$CDM cosmology with $\Omega_m = 0.308$, $\Omega_\Lambda=0.692$,  $\Omega_b=0.04$, $\sigma_8=0.829$, $n_s=0.961$ and  $H_0=100h$ km s$^{-1}$ cMpc$^{-1}$ = $67.8$  km s$^{-1}$ cMpc$^{-1}$ \cite{Planck2014}. The initial conditions for this simulation were generated at $z$ = 99 with the \texttt{N-GenIC} code, using the Zeldovich approximation, in a periodic cubic box of side length $160~\hcMpc$ containing $1024^3$ dark matter particles. The simulation outputs were saved at uniform redshift intervals of $\Delta z = 0.2$ from $z=20$ to $z=4.6$.

At each redshift, the discrete dark matter particle distribution was interpolated onto a three-dimensional grid with cell size $\Delta x_{\rm grid} = 2~h^{-1}\mathrm{cMpc}$, using the cloud-in-cell mass--assignment scheme, to obtain the corresponding dark matter density field $\rho_m(\boldsymbol{x})$. We further identified collapsed structures by employing the Friends-of-Friends (FoF) algorithm \cite{Davis1985_FoF} on the saved particle distribution. We used a minimum of ten particles as the criterion for identifying a halo, which set the smallest halo mass resolved in our simulations to be $3.2 \times 10^9~h^{-1} \Msun$ (= $10^{~9.67}~\Msun$). Our analysis throughout this paper is based on the \textbf{\emph{coeval}} simulation snapshot and halo catalogue saved at a fixed redshift of $z = 7$.

\subsection{Modeling UV Continuum and [O{\sc iii}] 5008$\angstrom$ Line Emission from High-$z$ Galaxies}
\label{subsec:galaxy_model}

Given a list of dark matter halos at the redshift of interest, we construct a corresponding galaxy catalogue by assigning high-redshift galaxies to these halos. Our approach builds on the methodology introduced in our earlier works \cite{Chakraborty2024, Chakraborty2025_Clustering, Chakraborty2025_DarkEnergy}, with additional improvements to better capture the galaxy-halo connection for massive halos.

For a given dark matter halo of mass $M_h$, we assign a stellar mass $M_\ast$ to its resident galaxy as
\begin{equation}
M_\ast(M_h) = f_\ast(M_h) \, \frac{\Omega_b}{\Omega_m} \, M_h
\label{eq:Mstar}
\end{equation}
where $f_\ast(M_h)$ is the star formation efficiency, defined as the fraction of baryons contained in the halo that are converted into stars.
The corresponding star formation rate (SFR) for the galaxy is then estimated as:
\begin{equation}
\dot{M}_\ast(M_h,z) =  \frac{M_\ast(M_h)}{t_\ast(z)}
\label{eq:sfr}
\end{equation}
where $t_\ast(z)$ denotes the characteristic timescale over which stars form at redshift $z$. 

We assume this timescale $t_\ast$ is proportional to the halo dynamical (or, equivalently, the free-fall) timescale, which in turn is proportional to the local Hubble timescale during the matter-domination era and can be expressed as -
\begin{equation}
t_\ast(z) = c_*~t_H(z)
\label{eq:tstar_defn}
\end{equation}
where $c_\ast$ is a dimensionless constant.

Given the SFR, one can directly compute the galaxy luminosity across different wavelengths. For instance, the rest-frame non-ionizing  UV luminosity at 1500 $\angstrom$ for a galaxy residing in a halo of mass $M_h$ is obtained using a mass--to--UV-light conversion factor $\mathcal{K}_{\mathrm{UV}}$, as follows:
\begin{equation}
L_{\mathrm{UV}}(M_h,z) = \frac{\dot{M}_\ast(M_h,z)}{\mathcal{K}_{\mathrm{UV}}}
\label{eq:luv_basic}
\end{equation}

From the above discussion, it is clear that several of our model parameters are mutually degenerate. For instance, the parameter $\kappaUV$  is fully degenerate with $f_\ast$ and $c_\ast$, since the UV luminosity of a galaxy depends only on the combination $f_\ast/(c_\ast\,\kappaUV)$. Consequently, different choices of individual parameters can lead to essentially identical predictions for the observables, and there may exist alternative parameter combinations that, in principle, provide equally good (or even better) fits to the available data. A systematic exploration of this multi-dimensional parameter space, and the associated degeneracies, lies beyond the scope of this proof-of-concept study. 

\begin{figure}[htbp]
\centering
\includegraphics[width=\columnwidth]{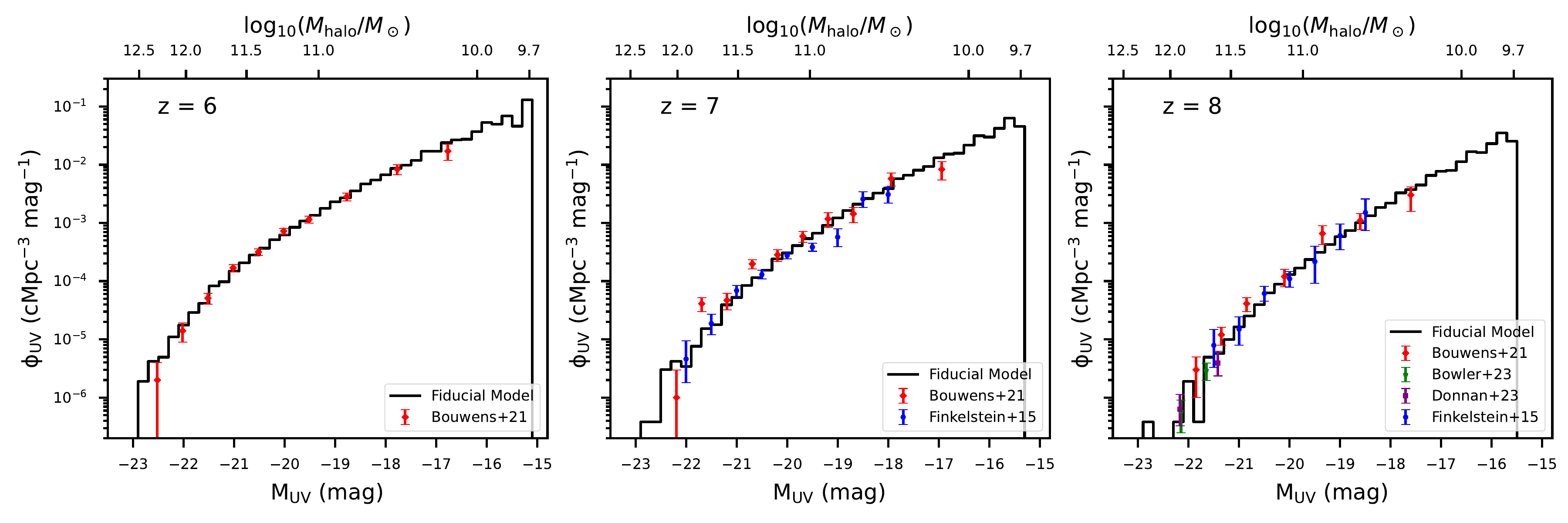}
\caption{The galaxy UV luminosity function (UVLF) $\Phi_{\rm UV}(\MUV,z)$ at $6 \leq z \leq 8$ as predicted by our fiducial model across the full simulation volume. In each panel, we also show various observational UVLF measurements \cite{Finkelstein2015, Bowler2020, Bouwens2021, Donnan2023} using colored data points. }
\label{fig:UVLF}
\end{figure}

In what follows, we therefore adopt a representative set of parameter values without loss of generality. This parameter set defines our \textbf{fiducial} model. For the star-formation timescale, we assume $t_\ast \simeq t_{\rm ff}$, which implies $c_\ast = 0.11$\footnote{The free-fall time of a virialized dark-matter halo, $t_{\rm ff} = t_{\rm dyn}/\sqrt{2} = \sqrt{3\pi/(32G\bar{\rho}_{\rm vir})}$, can be written in terms of the Hubble time as $t_{\rm ff}(z) = \pi/(2\sqrt{\Delta_c})\,t_H(z)$, assuming $\bar{\rho}_{\rm vir}(z) = \Delta_c\,\rho_c(z)$. For $\Delta_c \approx 18\pi^2$, this gives $t_{\rm ff}(z)\approx 0.11\,t_H(z)$.}. At $z=7$, this corresponds to $t_\ast \approx 125$ Myr, with a redshift evolution -- $t_\ast(z)\propto (1+z)^{-3/2}$. The mass-to-UV–light conversion factor is fixed to the commonly adopted value in the literature, following our previous work, $\kappaUVfid = 1.15\times10^{-28}\,\mathrm{M_\odot\,yr^{-1}}/(\mathrm{erg\,s^{-1}\,Hz^{-1}})$ \cite{MadauDickinson2014}. Finally, the star-formation efficiency $f_\ast$ of galaxies is modeled to be dependent on the mass of their host halo.  In this study, building on our previous works \cite{Chakraborty2024, Chakraborty2025_Clustering}, we update the prescription for $f_\ast$ by transitioning to a more flexible and physically-motivated double power-law functional form, which is given by 
\begin{equation}
f_\star(M_h) = 
\dfrac{2 f_{\star,12}}
{\left( \dfrac{M_h}{10^{12} M_\odot} \right)^{-\alpha_{\ast,\mathrm{lo}}}
 + \left( \dfrac{M_h}{10^{12} M_\odot} \right)^{-\alpha_{\ast,\mathrm{hi}}}},
\end{equation}
where $\alpha_{\ast,\mathrm{lo}}$ and $\alpha_{\ast,\mathrm{hi}}$ denote the power–law indices in the low– and high– halo mass regimes, respectively, and $f_{\star,12}$ sets the overall normalization by representing the star–formation efficiency at a characteristic mass-scale of $10^{12}\,M_\odot$. This widely adopted parameterization more faithfully captures the key mechanisms that regulate galaxy evolution across a wider range of halo masses -- for instance, stellar feedback in shallow potential wells efficiently depletes or heats the gas reservoir, suppressing star formation in low-mass halos, whereas the associated accelerated growth of accreting black-holes (active galactic nuclei) inhibits gas cooling and quenches star formation in the most massive halos. In our fiducial model, we adopt $\alpha_{\ast,\mathrm{lo}} = 0.3$, $\alpha_{\ast,\mathrm{hi}} = -0.61$, and $f_{\star,12}=0.036$.  The low–mass behaviour retains the scaling found in our previous works (e.g., \cite{Chakraborty2024, Chakraborty2025_Clustering}), while the high–mass behaviour is in agreement with other works in the literature (e.g., \cite{Mirocha2017, Davies2025}). As shown in \fig{fig:UVLF}, this \textbf{fiducial} model shows excellent agreement with the observational measurements of the UV luminosity function around the redshifts of interest to this study.

In practice, however, high-redshift galaxies selected using broadband photometry via their continuum emission (e.g., Lyman-break or other ``dropout'' techniques) typically have large redshift uncertainties ($\Delta z \gtrsim $ 0.1), corresponding to line-of-sight distance errors that severely compromise their utility for cross-correlation analyses with the 21\,cm signal (see, \cite{FurlanettoLidz2007, Lidz2009, LaPlante2023}). To circumvent this limitation, we instead focus on galaxies with spectroscopic redshifts, selected through the detection of nebular emission lines, which provide comparatively more precise estimates ($\Delta z \sim 0.01 - 0.001$) of galaxy redshifts. 

For our purposes, we focus on galaxies selected via their strong nebular $\OIII$ emission during the EoR. Such $\OIII$ line emitters are now routinely identified by deep near–infrared spectroscopic JWST surveys \cite{Jin2024, Meyer2024, Wold2025, Meyer2025, Korber2025}, and typically provide robust spectroscopic redshifts ($\Delta z \lesssim 0.01$). We stress, however, that there is nothing special about $\OIII$ emitters in the context of our methodology: the cross-correlation formalism we develop is fully generic and can be applied to any spectroscopically confirmed emission line-selected galaxy sample.

The $\OIIIang$ emission line is one of the strongest rest-frame optical nebular lines observed in the spectra of star-forming galaxies. It arises from a forbidden fine-structure transition ($^1D_2 \rightarrow {^3}P_2$) in doubly ionized oxygen (O$^{++}$) within H\textsc{ii} regions surrounding young, massive stars \cite{Osterbrock2006}. In these photoionized nebulae, O$^{++}$ ions are collisionally excited to metastable states and subsequently decay radiatively, producing the characteristic $\OIII$ doublet at 4959$\angstrom$ and 5008$\angstrom$. The strength of the $\OIII$ emission is governed primarily by two factors: (i) the rate at which hydrogen-ionizing photons are produced by the stellar population -- which, for a fixed stellar initial mass function and star-formation history, scales directly with the instantaneous SFR and (ii) the physical conditions of the interstellar gas, such as its metallicity, electron density, and ionization parameter \cite{Osterbrock2006, Kennicutt1998, Kewley2019}. Assuming photoionization equilibrium and negligible collisional de-excitation (as expected at the low gas densities characteristic of {H\sc{ii}} regions), the $\OIII$ luminosity can be taken to be proportional to the SFR, modulo secondary dependencies on the gas-phase metallicity (which sets the oxygen abundance and cooling efficiency) and ionization parameter (which determines the fraction of oxygen in the O$^{++}$ state).

Given the \textit{proof-of-concept} nature of this work, we do not attempt a full photoionization–based modelling of $\OIII$ line luminosities. Instead, motivated by the considerations above, we calculate the rest–frame $\OIIIang$ luminosity of a galaxy in our catalog from its star formation rate, as follows :
\begin{equation}
L_{\OIII}(M_h,z) = \CfacOIII \,\left[ 1 - f_{\mathrm{esc}}(M_h) \right]\, \dot{M}_\ast(M_h,z).
\label{eq:loiii_basic}
\end{equation}
where the parameter $\CfacOIII$ encapsulates the dependence of the $\OIII$ line luminosity on interstellar medium properties such as the gas-phase metallicity $Z$, density, and ionization parameter $U$. The factor $\left[1 - f_{\mathrm{esc}}(M_h)\right]$ accounts for the fact that only ionizing photons absorbed by neutral gas within the halo -- and thus, available to form and sustain H\textsc{ii} regions -- contribute to nebular line emission. 

Furthermore, in reality, high-redshift galaxy surveys typically cover a smaller comoving volume than that probed by the 21\,cm experiments. As a result, only the overlapping portion of the two surveys can contribute to the cross-correlation analysis.
To emulate this, we randomly extract a cubic sub-volume of $80^3 ~\cubedhcMpc$ from the full simulation box to serve as the galaxy survey region. 
In our case, this corresponds to a volume of $5.12\times 10^{5}~\cubedhcMpc$, comparable to that of the currently largest spectroscopic survey targeting $\OIII$ emitters during the reionization era  -- viz., JWST COSMOS-3D ($V_{\rm survey} \approx 6.3 \times 10^{5}~\cubedhcMpc ~\text{over}~6.75<z<7.5$) \cite{Meyer2025} -- and thus provides a realistic representation of the expected galaxy survey footprint. We note that extracting such a sub-volume automatically breaks the periodic boundary conditions of the parent simulation box. This loss of periodicity is \emph{unavoidable} in any realistic survey-like setup, where the observed galaxy field occupies a finite region embedded within a much larger cosmological volume.

To compute $L_{\OIII}(M_h,z)$ using \eqn{eq:loiii_basic}, we require a prescription for the escape fraction of ionizing photons, $f_{\mathrm{esc}}(M_h)$, and the parameter $\CfacOIII$. Consistent with our previous studies \cite{Chakraborty2024, Chakraborty2025_Clustering}, we model the ionizing escape fraction as a power-law function of the host halo mass, given by
\begin{equation}
f_\mathrm{esc}(M_h) = f_{\mathrm{esc,10}}\left( \dfrac{M_h}{10^{10} M_\odot} \right)^{\alpha_{\mathrm{esc}}}
\label{eq:powerlaw_fesc}
\end{equation}
where $f_{\mathrm{esc,10}}$ is the normalization at $M_h = 10^{10}\,M_\odot$ and $\alpha_{\mathrm{esc}}$ controls the mass dependence.

\begin{figure}
\centering
\includegraphics[width=0.9\columnwidth]{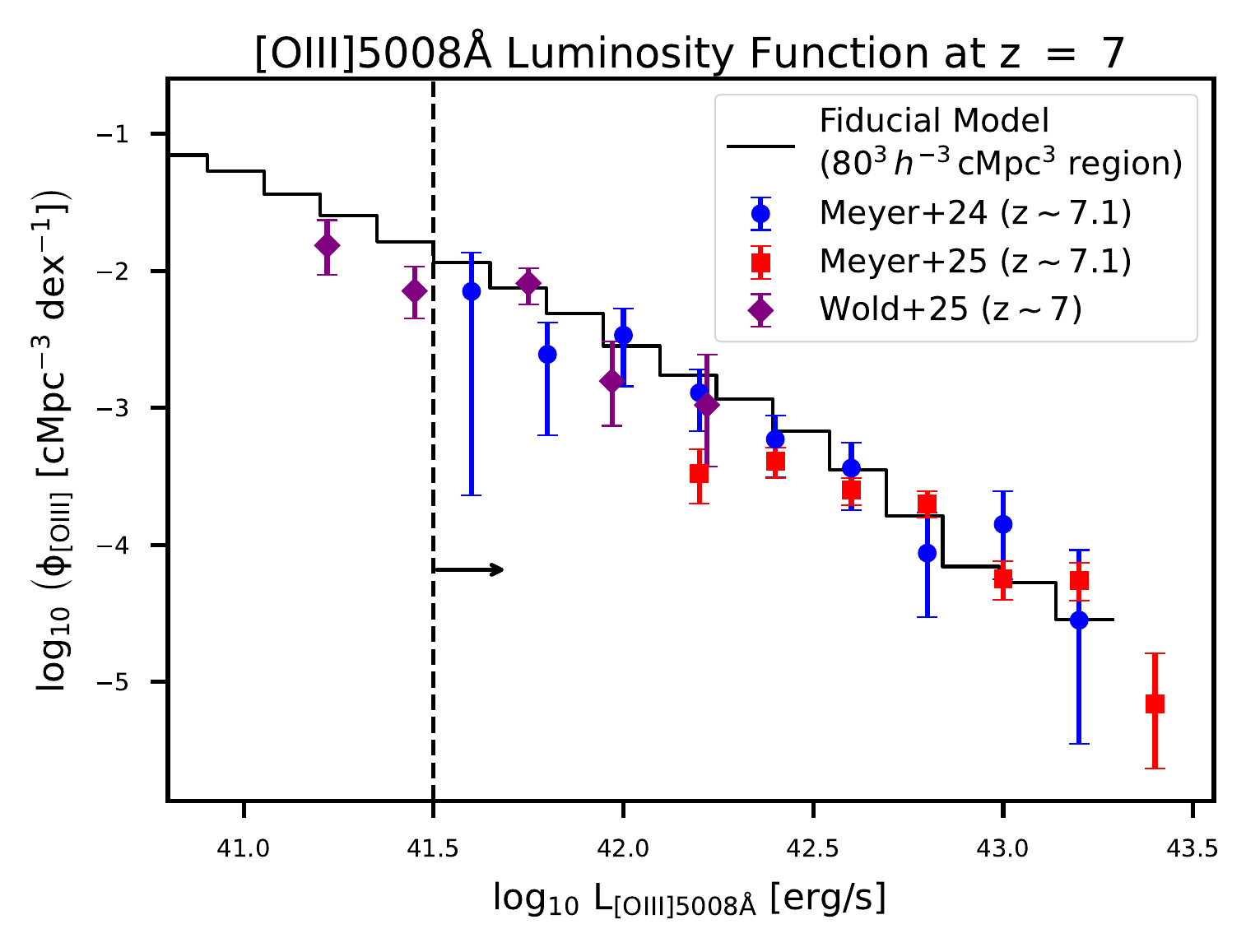}
\caption{The galaxy $\OIIIang$ luminosity function (O3LF) at $z = 7.0 $ predicted by our fiducial model for the randomly selected survey region having a volume of $80^3~\cubedhcMpc$. We also show various observational measurements at $z\sim 7.1$ from various JWST surveys 
\cite{Wold2025, Meyer2024, Meyer2025} using colored data points.}
\label{fig:O3LF}
\end{figure}
For the \textbf{fiducial} model, we adopt $f_{\mathrm{esc,10}} = 0.09$, $\alpha_{\mathrm{esc}} = -0.3$, following our previous work \cite{Chakraborty2025_Clustering}. As we will discuss in the next subsection, these parameter choices lead to a reionization history that is consistent with current observational constraints.  The value of $\CfacOIII$ is obtained by calibrating the resulting $\OIII$ luminosity function from our surveyed region to reproduce observational
measurements. At $z \sim 7$, we find that
$\CfacOIII = 4.25 \times 10^{41}\,\mathrm{erg\,s^{-1}}/(\mathrm{M_\odot\,yr^{-1}})$ produces a good match, as shown in \fig{fig:O3LF}.

While constructing our mock sample of $\OIII$ emitters, we further impose a luminosity threshold of
$L_{\OIII} \geq 10^{41.5}\,\mathrm{erg\,s^{-1}}$, chosen to be broadly representative of the typical $\OIII$ detection sensitivity achieved by current blank-field JWST surveys. Although fainter $\OIII$-emitting sources have been detected, they are predominantly identified
in surveys of strongly lensed fields (e.g., Abell~2744 HFF \cite{Wold2025}; Abell~S1063 HFF \cite{Korber2025}), and estimates of
their number densities and observed $\OIII$ luminosities are subject to additional uncertainties arising from lensing magnification corrections. Applying this luminosity threshold results in a total of $N_{\rm gal}= 7456$ galaxies in our mock catalog.

\subsection{Modeling the Fluctuating 21\,cm Signal from the Intergalactic Medium}
\label{subsec:reion_21cm_model}

Having discussed the emission properties of the high-redshift galaxy population, we now describe the modelling of their ionizing photon output and the construction of maps of the 21\,cm emission from intergalactic neutral hydrogen (H\textsc{i}) in this subsection.

We assume that the intrinsic number of ionizing photons produced by a halo is proportional to its stellar mass. Under this assumption, the number of ionizing photons that a halo of mass $M_h$ leaks into the IGM can be written as
\begin{align}
N_{\mathrm{ion}}(M_h) &= f_{\mathrm{esc}}(M_h) \, N_{\mathrm{ion,int}}(M_h) \nonumber \\
&= f_{\mathrm{esc}}(M_h) \, \eta_\gamma \, M_\ast(M_h)
\label{eq:nion}
\end{align}
where $\eta_\gamma$ is the number of ionizing photons produced intrinsically for every solar mass of stars formed and $f_{\mathrm{esc}}(M_h)$ is the fraction of these photons that escape into the IGM from the galaxy.

Using the stellar-to-halo mass relation given in \eqn{eq:Mstar}, we can simplify \eqn{eq:nion} further :
\begin{align}
N_{\mathrm{ion}}(M_h) &= f_{\mathrm{esc}}(M_h) \, \eta_\gamma \, f_\ast(M_h) \,  \frac{\Omega_b}{\Omega_m} \, M_h \\
&= \zeta(M_h) \, \frac{1 - Y}{m_{\mathrm{p}}} \, \frac{\Omega_b}{\Omega_m} \, M_h
\label{eq:nion_2}
\end{align}
where $\zeta(M_h)$ denotes the ionizing efficiency parameter (i.e., the number of ionizing photons emitted per hydrogen atom) and is defined as 
\begin{equation}
\zeta(M_h) = \frac{f_{\mathrm{esc}}(M_h) \, f_\ast(M_h)  \, \eta_\gamma 
 \,  m_{\mathrm{p}}}{1 - Y} 
\label{eq:zeta_defn}
\end{equation}

Therefore, for a particular grid cell $i$, the comoving number density of ionizing photons contributed by halos with masses greater than the atomic cooling threshold $M_{\rm cool}$ can be expressed in terms of the conditional halo mass function in that cell as 
\begin{align}
n_{\mathrm{ion},i} 
    &= \int_{M_{\mathrm{cool}}}^\infty \mathrm{d}M_h \, \frac{\mathrm{d}n}{\mathrm{d}M_h}\bigg|_i \, N_{\mathrm{ion},i}(M_h) \nonumber \\
    &= \frac{1 - Y}{m_{\mathrm{p}}} \, \frac{\Omega_b}{\Omega_m} \int_{M_{\mathrm{cool}}}^\infty \mathrm{d}M_h \, \frac{\mathrm{d}n}{\mathrm{d}M_h}\bigg|_i \, \zeta(M_h) \, M_h \nonumber \\[6pt]
    &= n_{\mathrm{H},i} \, \bigg[\zeta \, f_{\mathrm{coll}}(M_h \geq M_\mathrm{cool}) \bigg]_i
    \label{eq:nion_cell_final}
\end{align}
where $n_{\mathrm{H},i}$ is the comoving hydrogen number density in the $i^{\rm th}$ grid cell, and
\begin{equation*}
\bigg[\zeta \, f_{\mathrm{coll}}(M_h \geq M_\mathrm{cool}) \bigg]_i \;\equiv\;
\frac{1}{\rho_{m,i}}
\int_{M_{\mathrm{cool}}}^{\infty} \mathrm{d}M_h \,
\frac{\mathrm{d}n}{\mathrm{d}M_h}\bigg|_i \,
\zeta(M_h)\, M_h ~,
\end{equation*}
represents the ionizing-efficiency–weighted mass fraction in collapsed halos of mass $M_h \ge M_{\mathrm{cool}}$.

At the redshift of our interest ($z = 7$), dark matter halos down to the atomic cooling threshold (i.e., 
$M_\mathrm{cool} \approx 10^{~8.13}\, M_\odot$) are not resolved in our $N$-body 
simulations. In fact, the smallest halos identified from our $N$-body simuation are comparatively more massive ($M^{\rm sim}_{h,\mathrm{min}} \approx 10^{~9.67}\, M_\odot$). Nevertheless, as evident 
from \eqn{eq:nion_cell_final}, the comoving ionizing photon number density in each grid cell, $n_{\mathrm{ion},i}$, can be cleanly separated into two components: one arising 
from halos explicitly identified by the group finder and another from unresolved, 
sub-grid halos that lie below the simulation resolution limit : 
\begin{align}
n_{\mathrm{ion},i} &= n_{\mathrm{H},i}\, \bigg[\zeta\, f_{\mathrm{coll}}(M_h \geq M_\mathrm{cool})\bigg]_i 
\nonumber \\[6pt]
&= n_{\mathrm{H},i}\, \bigg[\zeta\, f_{\mathrm{coll}}(M_\mathrm{cool} \leq M_h < 
M^{\rm sim}_{h,\mathrm{min}}) + \zeta\, f_{\mathrm{coll}}(M_h \geq 
M^{\rm sim}_{h,\mathrm{min}})\bigg]_i 
\label{eq:nion_cell_final_split}
\end{align}

In the above expression, the first term captures the contribution from sub-grid halos and is computed using a semi-analytical prescription for $\mathrm{d}n/\mathrm{d}M_h\big|_i$, based on the conditional ellipsoidal collapse of density perturbations described in Choudhury \& Paranjape (2018)\cite{Choudhury2018}. The second term, corresponding to halos explicitly identified in the simulation, is calculated directly from the halo catalog as  --
\begin{equation}
\label{zeta_fcoll_eqn}
\bigg[\zeta\, f_{\mathrm{coll}}(M_h \geq M^{\rm sim}_{h,\mathrm{min}})\bigg]_i
= \dfrac{ \displaystyle  \sum_{\substack{m_h \in  i^{\mathrm{th}}\text{cell}}}\zeta(m_h)\, m_h}
    {\displaystyle M_{\mathrm{total},i}}
\end{equation}
where $M_{\mathrm{total},i}$ is the total dark matter mass contained in the 
$i^{\rm th}$ grid cell and $m_h$ is the mass of an individual halo within that cell.

We use the semi-numerical code \texttt{SCRIPT} (\textbf{S}emi-numerical \textbf{C}ode for \textbf{R}e\textbf{I}onization with \textbf{P}ho\textbf{T}on-conservation) to generate the ionization fields \cite{Choudhury2018}. The photon-conserving algorithm implemented in \texttt{SCRIPT} proceeds in two stages. In the first stage, ionized bubbles are generated around individual sources within the simulation box. For a given grid cell $i$ that produces $N_{\mathrm{ion},i} = n_{\mathrm{ion},i} ~ (\Delta x_{\mathrm{grid}})^3$
ionizing photons, we first consume $N_{\mathrm{H},i} = n_{\mathrm{H},i} ~ (\Delta x_{\mathrm{grid}})^3$ of the $N_{\mathrm{ion},i}$ photons to ionize the hydrogen atoms in the source cell itself. Any remaining photons are then distributed to surrounding cells in order of increasing distance from the source, until all the photons produced in that cell are exhausted. 

After accounting for all photons incident on a given grid cell $i$, if the total number of \emph{available} ionizing photons $N^{\rm avail}_{\mathrm{ion},i}$ is found to exceed the number of hydrogen atoms $N_{\mathrm{H},i}$ present, then that cell is marked as fully ionized ($x_{\mathrm{HII},i} = 1$), with the surplus ionizing
photons retained for subsequent redistribution. If instead
$N^{\rm avail}_{\mathrm{ion},i} \leq N_{\mathrm{H},i}$, the particular cell is assigned a
ionized fraction equal to
\begin{equation}
x_{\mathrm{HII},i} = \dfrac{N^{\rm avail}_{\mathrm{ion},i}}{N_{\mathrm{H},i}}
\end{equation}
This procedure is carried out independently for all source cells in the box.  As a consequence, some grid cells may receive ionizing photons from multiple sources and can become ``over-ionized''.
Thus, in the second stage of the algorithm, the excess ionizing photons in these ``over-ionized'' cells are again redistributed to neighbouring cells that are not yet fully ionized. This redistribution is carried out iteratively until all ``over-ionized'' cells are properly accounted for. 

This approach offers important advantages over other existing semi-numerical methods of simulating cosmic reionization. Most notably, it enforces explicit conservation of ionizing photon number, thereby resolving a well-known limitation of earlier excursion-set–based approaches (see, e.g., \cite{Zahn2007, Paranjape2016, Hutter2018}). Consequently, it also ensures the numerical convergence of large-scale power spectra of the ionization fluctuations with respect to the resolution of the ionization maps \cite{Choudhury2018, Maity2023}. In summary, given a prescription for the ionizing efficiency of halos, $\zeta(M_h)$, along with the gridded matter overdensity field $\Delta_m(\boldsymbol{x}) \equiv \rho_m(\boldsymbol{x})/\bar{\rho}_m $ and the masses and spatial positions of dark matter halos identified in a $N$-body simulation, \texttt{SCRIPT} computes and outputs the ionized hydrogen fraction $x_{\mathrm{HII},i}$ for each grid cell of the simulation volume.

\begin{figure}[htbp]
\centering
\includegraphics[width=0.7\columnwidth]{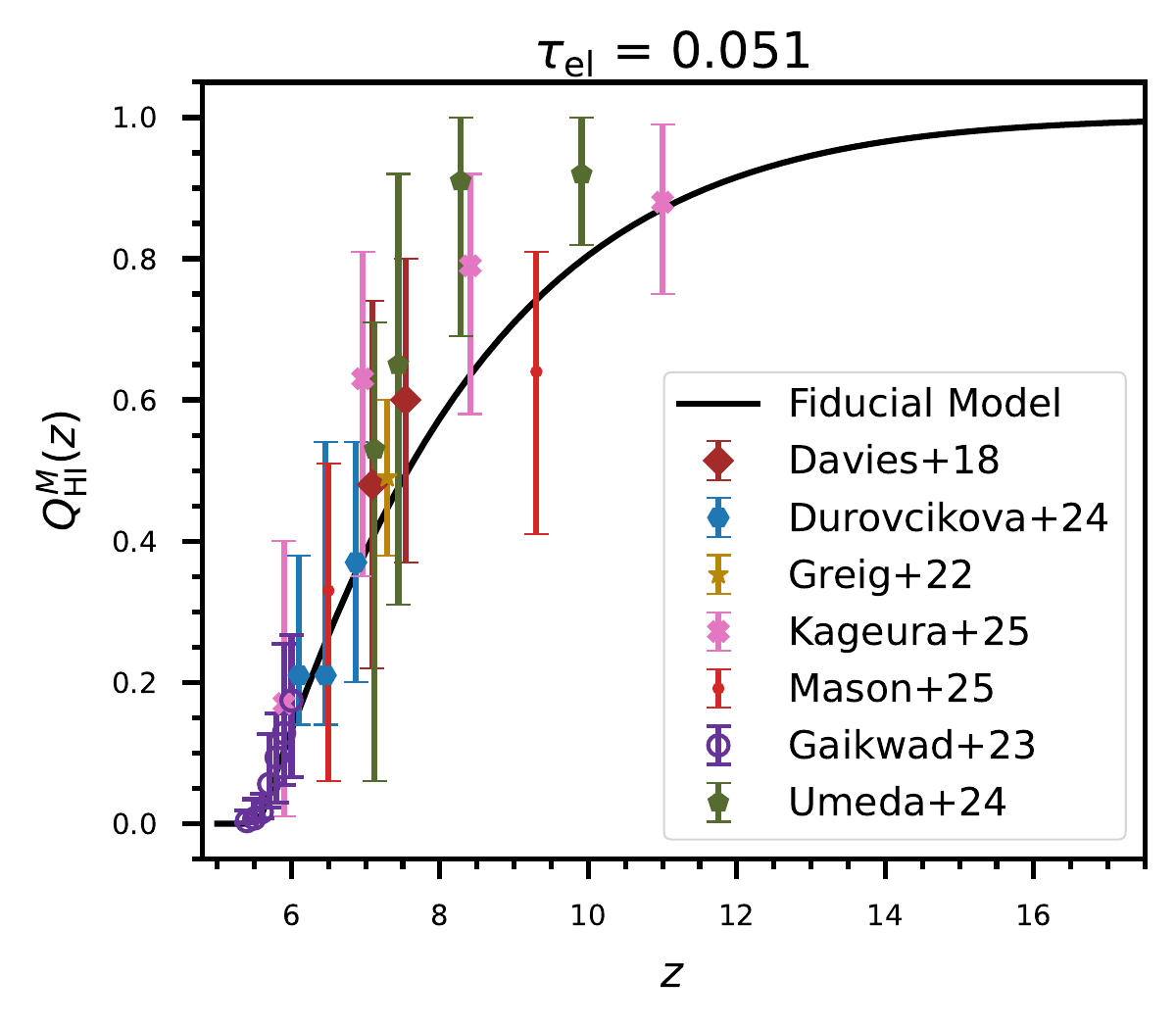}
\caption{The redshift evolution of the mass-weighted neutral hydrogen fraction predicted by our \textbf{fiducial} model. At redshift $z = 7$, this model corresponds to a mean mass-weighted neutral fraction of $Q^{M}_{\mathrm{HI}}\equiv \langle(1-x_{\mathrm{HII},i})~\Delta_i
\rangle$ = 0.385, where the average is taken over the full simulation volume.  The colored data points
represent some of the latest observational measurements \cite{Davies2018,Durovcikova2024,Greig2022,Kageura2025,Mason2025,Gaikwad2023,Umeda2023}.}
\label{fig:reionhist}
\end{figure}

For the \textbf{fiducial} model, the parameters describing the ionizing escape fraction $f_{\rm esc}(M_h)$ of
halos (see \eqn{eq:powerlaw_fesc}) are fixed by requiring the resulting reionization history to produce a value of the
Thomson scattering optical depth of CMB photons, $\tau_{\rm el}$, consistent with the
latest measurements from the Planck Collaboration \cite{Planck2018}, while also satisfying
existing observational constraints on the redshift evolution of the globally averaged neutral hydrogen fraction,
$Q_{\mathrm{HI}}(z)$. We adopt $\eta_\gamma m_{\mathrm{p}} = 3990$, similar to our previous works \cite{Chakraborty2025_Clustering} and assume a $f_{\mathrm{esc,10}} = 0.09$, $\alpha_{\mathrm{esc}} = -0.3$. This choice results in an optical depth of $\tau_{\rm el} = 0.051$, with reionization completing at $z \approx 5.6$, as shown in \fig{fig:reionhist}. 

\begin{figure}[htbp]
\centering
\includegraphics[width=0.8\columnwidth]{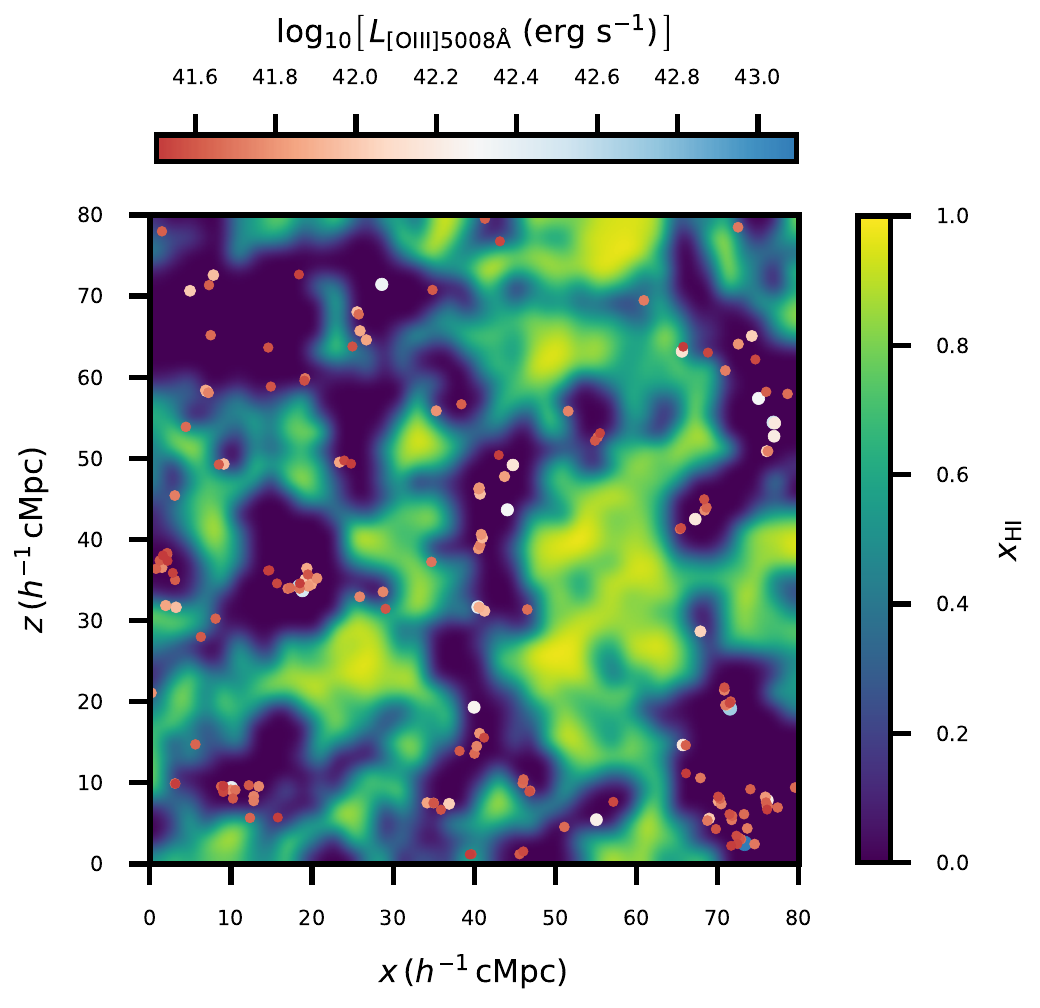}
\caption{A two-dimensional slice of thickness $2\,h^{-1}\,\mathrm{cMpc}$ through the galaxy survey region, showing the spatial distribution of $\OIII$ emitters brighter than $10^{41.5}\,\mathrm{erg\,s^{-1}}$ (shown as points and coloured by their $\OIII$ luminosity), overlaid on the neutral hydrogen fraction field, $x_{\mathrm{HI},i} \equiv 1 - x_{\mathrm{HII},i}$.}
\label{fig:2Dslice_gal_21cm}
\end{figure}

\fig{fig:2Dslice_gal_21cm} shows a two-dimensional visualization of the galaxy–IGM connection from our simulations, depicting the spatial locations of bright $\OIII$ emitters and the underlying neutral hydrogen distribution across the selected galaxy survey volume. As seen in the figure, galaxies preferentially reside within ionized regions (dark-colored patches).

To connect the ionization field to an observable quantity, we compute the corresponding 21 cm emission signal from neutral hydrogen, which is measured as a brightness temperature relative to the cosmic microwave background (CMB). Assuming that the gas spin temperature is much larger than the CMB temperature, the differential 21\,cm brightness temperature $\delta T_{21,i}$ in each grid cell is given by
\begin{equation}
\label{eq:delta_Tb}
\delta T_{21, i} \approx 27~\mathrm{mK}
\left(1 - x_{\mathrm{HII}, i}\right) \Delta_i
\left(\frac{1+z}{10}\frac{0.15}{\Omega_{m}h^2}\right)^{1/2}
\left(\frac{\Omega_{b}h^2}{0.023}\right),
\end{equation}
where $\Delta_i \equiv \rho_{m,i}/\bar{\rho}_m$  is the ratio of the matter density $\rho_{m,i}$ in the grid cell $i$ and the mean cosmic matter density $\bar{\rho}_{m}$.

For our analysis, to more realistically model interferometric 21\,cm observations, we further incorporate a range of observational effects into the simulated $\delta T_{21}$ field obtained from \texttt{SCRIPT}. Specifically, we account for the contamination of the cosmological 21\,cm signal from instrumental thermal noise and astrophysical foregrounds, as outlined below:

\begin{itemize}
    \item \textbf{Contamination from Instrumental Thermal Noise}

    To the simulated $\delta T_{21}$ cubes, we add thermal noise assuming an observing campaign with the SKA-Low telescope in AA$^\ast$ configuration, which consists of 307 antenna elements and corresponds to the layout planned for early science observations.
    Our campaign targets a sky region that transits the zenith at a declination of $-30^\circ$ and is assumed to be conducted for 6 hours per day over 20 days, resulting in a total on-source integration time of 120 hours.  Along the line-of-sight direction, the comoving length of our coeval simulation box ($L_{\rm box} = 160 \, \hcMpc$) at $z = 7$ corresponds to a total observing bandwidth of $B_\nu \approx 14.892$ MHz, which is divided into 80 discrete frequency channels, each with a width of $ \Delta \nu_{\rm ch} \simeq 0.186$ MHz.

    Using these specifications, we generate the daily $uv$-coverage tracks and, thereafter, Gaussian random realizations of the instrumental thermal noise, using the publicly available \texttt{tools21cm}\footnote{https://github.com/sambit-giri/tools21cm}
     package \cite{giri2020tools21cm}, following the methodology described in Giri et al. (2018) \cite{Giri2018}. The noise cubes are produced at a spatial resolution matched to that of the \texttt{SCRIPT} simulations. As a result, the angular resolution of the resulting 21\,cm images at $z = 7$ is approximately $1.15$ arcminutes, corresponding to a comoving length scale of $\Delta x_{\rm grid} = 2~\hcMpc$. We emphasize that throughout this analysis, we neglect any evolution of the 21\,cm signal and thermal noise along the frequency (line-of-sight) axis and therefore employ coeval cubes at $z=7$ for both.

    \item \textbf{Contamination from Astrophysical Foreground Emission}  
    
    Bright astrophysical foregrounds in the form of Galactic synchrotron emission and emission of extragalactic point sources are typically several orders of magnitude stronger than the cosmological 21\,cm signal. Although astrophysical foregrounds are intrinsically spectrally smooth, the frequency-dependent response of radio interferometric telescopes introduces chromatic effects that mix transverse ($k_\perp$) and line-of-sight ($k_\parallel$) modes during the reconstruction of the 21\,cm brightness temperature fluctuations, $\delta T_{21}(\boldsymbol{x})$, from the measured visibilities. As a result, foreground power that would otherwise be confined to low line-of-sight wavenumbers leaks into higher $k_\parallel$ modes. This leakage produces a characteristic wedge-shaped region in $(k_\perp, k_\parallel)$ space, commonly referred to as the foreground wedge. Fourier modes within this region are expected to be strongly contaminated by foreground emission and are therefore difficult to model or remove reliably in practice.

    To mitigate contamination from bright astrophysical foregrounds, we implement a foreground avoidance strategy by discarding all Fourier modes that lie below the horizon limit of the foreground wedge. Specifically, we first transform the noise-contaminated 21\,cm fields into the three-dimensional Fourier space, then explicitly set the amplitude of all Fourier modes that fall within the wedge region to zero, and finally apply an inverse Fourier transform to obtain the foreground-filtered fields in real space.
    The boundary of this wedge in two-dimensional Fourier space is usually defined as
    \begin{equation}
        k_{\parallel} \leq \sin\theta~\dfrac{D_c(z) H(z)}{c (1 + z)}~k_{\perp},
        \label{eq:wedge_equation}
    \end{equation}
    where $k_\perp$ and $k_\parallel$ denote the components of the wavevector $\boldsymbol{k}$ perpendicular and parallel to the line of sight (taken to be along the $z$-axis of the simulation box), respectively, $D_c(z)$ is the comoving distance to redshift $z$, and $\theta$ is the angular radius of the interferometer beam. To maximize the impact of foreground avoidance, we adopt a pessimistic assumption of $\theta = 90^\circ$, which evaluates to $k_{\parallel} \lesssim 3.14\,k_{\perp}$. This choice is motivated by the realistic response of antenna beams, which typically taper off smoothly rather than terminating abruptly. As a result, sidelobe responses can also pick up strong foreground emission from regions far away from the zenith.
\end{itemize}

Finally, we note that in our analysis, we work with mean-subtracted 21\,cm brightness temperature fields to account for the fact that radio interferometers lack zero-spacing baselines and are therefore insensitive to the global (spatially-averaged) 21\,cm signal.

\begin{figure}
    \centering
    \includegraphics[width=\textwidth]{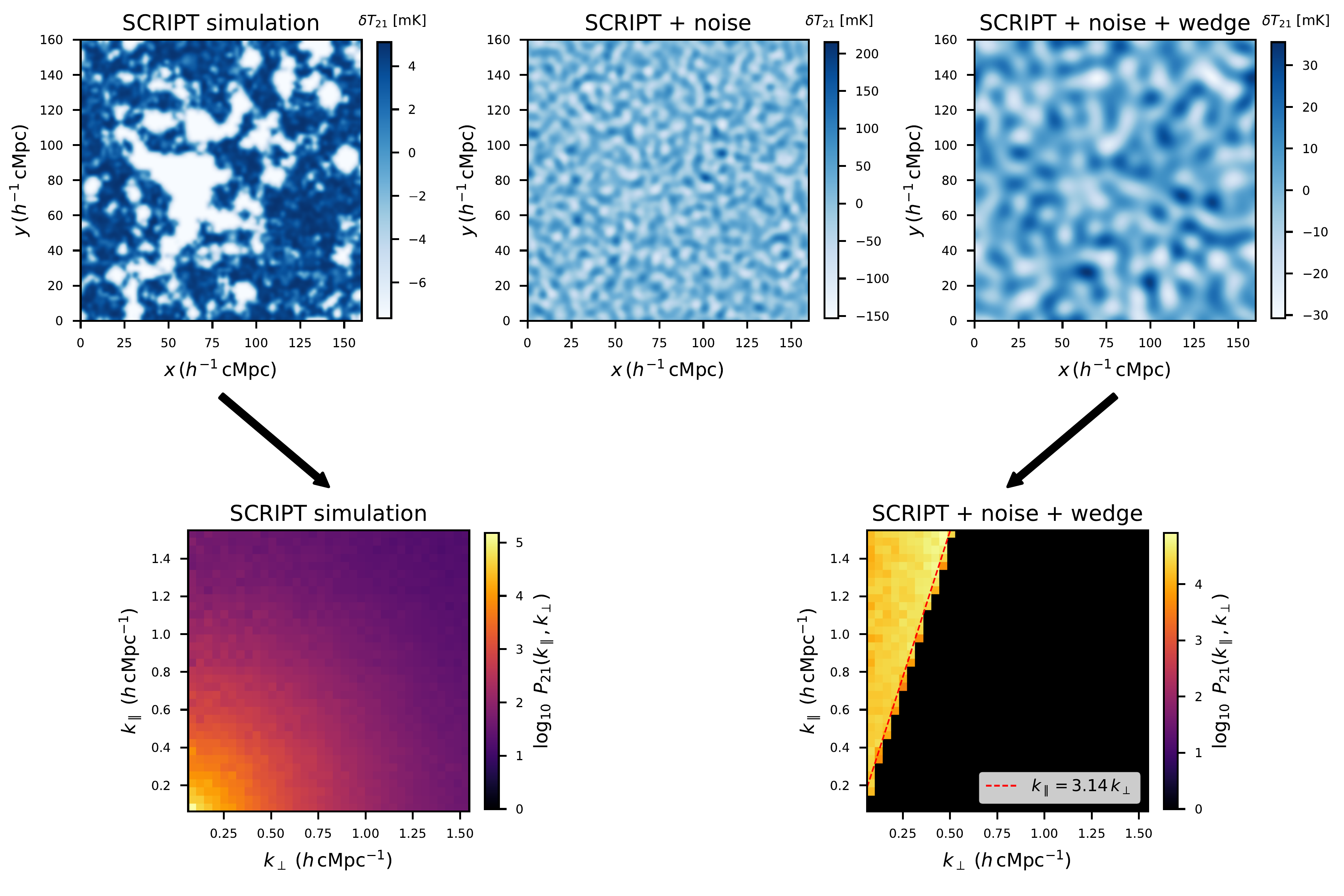}
    \caption{
    Effect of instrumental noise and foreground wedge filtering on real-space 21\,cm maps and the power spectra of 21\,cm fluctuations.\\
    \textbf{\emph{Top row:}} Two-dimensional slices of the mean-subtracted 21\,cm brightness temperature fluctuation field,
    $\delta T_{21}$, from the \texttt{SCRIPT} simulation (left), after adding thermal noise (middle),
    and after applying both thermal noise and foreground wedge filtering (right).
    The colour bars show the value of $\delta T_{21}$ in units of mK. \textbf{\emph{Bottom row:}} The binned cylindrically averaged power spectra,
    $\log_{10} P_{21}(k_\parallel, k_\perp)$, of fluctuations in $\delta T_{21}$.
    In the right-most panel, the red dashed line marks the foreground wedge boundary,
    $k_\parallel = 3.14\,k_\perp$. }
    \label{fig:script_noise_wedge_illustration}
\end{figure}

In \fig{fig:script_noise_wedge_illustration}, we illustrate how these effects progressively modify the simulated 21\,cm signal. The noiseless 21\,cm fields produced by \texttt{SCRIPT} exhibit coherent large-scale structure in real space and a smooth distribution of power across the cylindrical Fourier space, with most of the power concentrated at low $k_\perp$ and low $k_\parallel$. The addition of thermal noise introduces significant small-scale fluctuations that, in real space, dominate over the underlying cosmological signal, as shown in the top-middle panel. The subsequent application of foreground wedge filtering suppresses a substantial fraction of these fluctuations by removing contaminated Fourier modes, leaving only modes outside the wedge boundary accessible for analysis; nevertheless, the residual fluctuations remain larger than those of the underlying cosmological signal (see top-left and top-right panels).

For the 21\,cm--galaxy cross-correlation analysis, an additional consideration is the mismatch between the volumes probed by the two observations. As discussed earlier, planned SKA-Low 21\,cm experiments will measure the 21\,cm signal over comoving volumes that are substantially larger than those accessible to existing high-redshift galaxy surveys, spanning hundreds to thousands of square degrees on the sky. Accordingly, we simulate the 21\,cm signal over the entire $160^3~ \cubedhcMpc$ volume, and subsequently, apply instrumental noise and foreground-wedge filtering to this volume. We then extract an appropriately sized 21\,cm sub-volume, centered on the galaxy survey region, from the simulated $160^3~\cubedhcMpc$ volume so that the 21\,cm and galaxy fields share a common survey footprint for the cross-correlation analysis.


\section{Frameworks for Computing Cross-Correlations}
\label{sec:crosscorr_frameworks}

In this section, we outline the statistical frameworks used to quantify spatial cross-correlations between discrete tracers and a continuous field.

\subsection{Two-point Cross Correlation Functions}
\label{subsec:crosscorr_2ptstacking}

The two-point cross-correlation function provides the simplest and most commonly used measure of spatial cross-correlations between two datasets. We consider a continuous field $\rho(\boldsymbol{x})$ defined in a cubic box of side length $L$, with zero mean, $\langle \rho(\boldsymbol{x}) \rangle = 0$. 
In our case, this field corresponds to the mean-subtracted 21\,cm brightness temperature fluctuation field -- i.e., $\rho(\boldsymbol{x}) \equiv \delta T_{21}(\boldsymbol{x}) - \langle \delta T_{21}(\boldsymbol{x}) \rangle$.
Let $\{\boldsymbol{x}_{\mathrm{gal},i}\}$ denote the positions of $N$ discrete tracers/galaxies, each labelled by $i$, within the same volume $V = L^3$. The distribution of discrete tracers can also be equivalently represented by a dimensionless continuous field $g(\boldsymbol{x})$ that describes fluctuations in the number density,
\begin{equation}
g(\boldsymbol{x})
=
\frac{
\displaystyle \sum_{i=1}^{N}
\delta_{\rm D}\!\left(\boldsymbol{x}-\boldsymbol{x}_{{\rm gal},i}\right)
}{
\bar{n}
} - 1 
\label{eq:galaxy_field} 
\end{equation}
Here, $\delta_{\rm D}$ denotes the Dirac-Delta function and
\(\bar{n} = N/V\) is the mean galaxy number density within the
volume $V$.

The two-point galaxy--field cross-correlation function at a length scale $r$ is defined as
\begin{equation}
\xi_{\mathrm{cross}}(r)
\equiv
\bigg\langle
g(\boldsymbol{x})\, \rho(\boldsymbol{x}')
\bigg\rangle_{|\boldsymbol{x}-\boldsymbol{x}'|=r}
\end{equation}
where the angular brackets denote an average over all pairs of points
$(\boldsymbol{x},\boldsymbol{x}')$ satisfying
$|\boldsymbol{x}-\boldsymbol{x}'|=r$. This average can be written out explicitly as follows,
\begin{align}
\xi_{\mathrm{cross}}(r)
&=
\frac{1}{V}
\int_V {\rm d}^3\boldsymbol{x}\, g(\boldsymbol{x})\,
\left[\rho(\boldsymbol{x}')
\right]_{|\boldsymbol{x}'-\boldsymbol{x}|=r}
 \nonumber \\ 
&=\frac{1}{V}
\int_V {\rm d}^3\boldsymbol{x}\, g(\boldsymbol{x}) \,
\frac{1}{V}
\int_V {\rm d}^3\boldsymbol{x}'\,
\rho(\boldsymbol{x}')\,
\delta_{\rm D}\!\left(|\boldsymbol{x}'-\boldsymbol{x}|-r\right)
\end{align}
where in the second line we have enforced the fixed separation $r$ by restricting the calculation to only those points
$\boldsymbol{x}'$ that lie at a radial distance $r$ from $\boldsymbol{x}$, using a Dirac delta function.

Substituting the form of $g(\boldsymbol{x})$ from \eqn{eq:galaxy_field} and
using the fact that $\langle \rho(\boldsymbol{x}) \rangle = 0$, we obtain
\begin{align}
\xi_{\mathrm{cross}}(r)
&=
\frac{1}{V}
\int_V \mathrm{d}^3 \boldsymbol{x}'\,\rho(\boldsymbol{x}')\,
\frac{1}{V\, \bar{n}}\int_V \mathrm{d}^3 \boldsymbol{x}\,
\sum_{i=1}^{N}
\delta_{\rm D}\!\left(\boldsymbol{x}-\boldsymbol{x}_{{\rm gal},i}\right)
\,
\delta_{\rm D}\!\left(|\boldsymbol{x}'-\boldsymbol{x}|-r\right)
\nonumber \\
&=
\frac{1}{N}
\sum_{i=1}^{N}
\frac{1}{V}
\int_V {\rm d}^3\boldsymbol{x}'\,
\rho(\boldsymbol{x}')\,
\delta_{\rm D}\!\left(|\boldsymbol{x}'-\boldsymbol{x}_{{\rm gal},i}|-r\right).
\end{align}

In practice, the Dirac delta function is implemented using a spherical top-hat filter of radius $r$ and width $\mathrm{d}r$, resulting in the following estimator.
\begin{equation}
\hat{\xi}_{\rm cross}(r)
=
\frac{1}{N}
\sum_{i=1}^{N}
\frac{1}{V_{\rm shell}(r)}
\int_{r \le |\boldsymbol{x}'-\boldsymbol{x}_{{\rm gal},i}| < r+\mathrm{d}r}
\rho(\boldsymbol{x}')\, {\rm d}^3\boldsymbol{x}' ,
\end{equation}
where
\begin{equation}
V_{\rm shell}(r)
=
\frac{4\pi}{3}\left[(r+\mathrm{d}r)^3 - r^3\right].
\end{equation}

This implies that formally, the two-point cross correlation function $\xi_{\mathrm{cross}}(r)$ for a particular length scale $r$ corresponds to the mean value of the continuous field
$\rho(\boldsymbol{x})$, averaged over spherical shells of radius $r$ and thickness
$\mathrm{d}r$, around the positions of all galaxies in the total sample.

To compute the two-point cross-correlation function $\xi_{\rm cross}(r)$ between the discrete galaxies and the continuous 21\,cm field over the range: $4~\hcMpc \leq r \leq 12~\hcMpc$, we employ this stacking procedure, as also described in Banerjee \& Abel (2023) \cite{Banerjee2023_tracerfieldCross}.
For a given scale $r$, we define spherical shells of radius $r$ and thickness $1 \hcMpc$ centered on each galaxy and average the 21\,cm field within these shells over all 7456 galaxies in our sample to obtain the cross-correlation at that length scale.

\begin{figure}
    \centering
    \includegraphics[width=\textwidth]{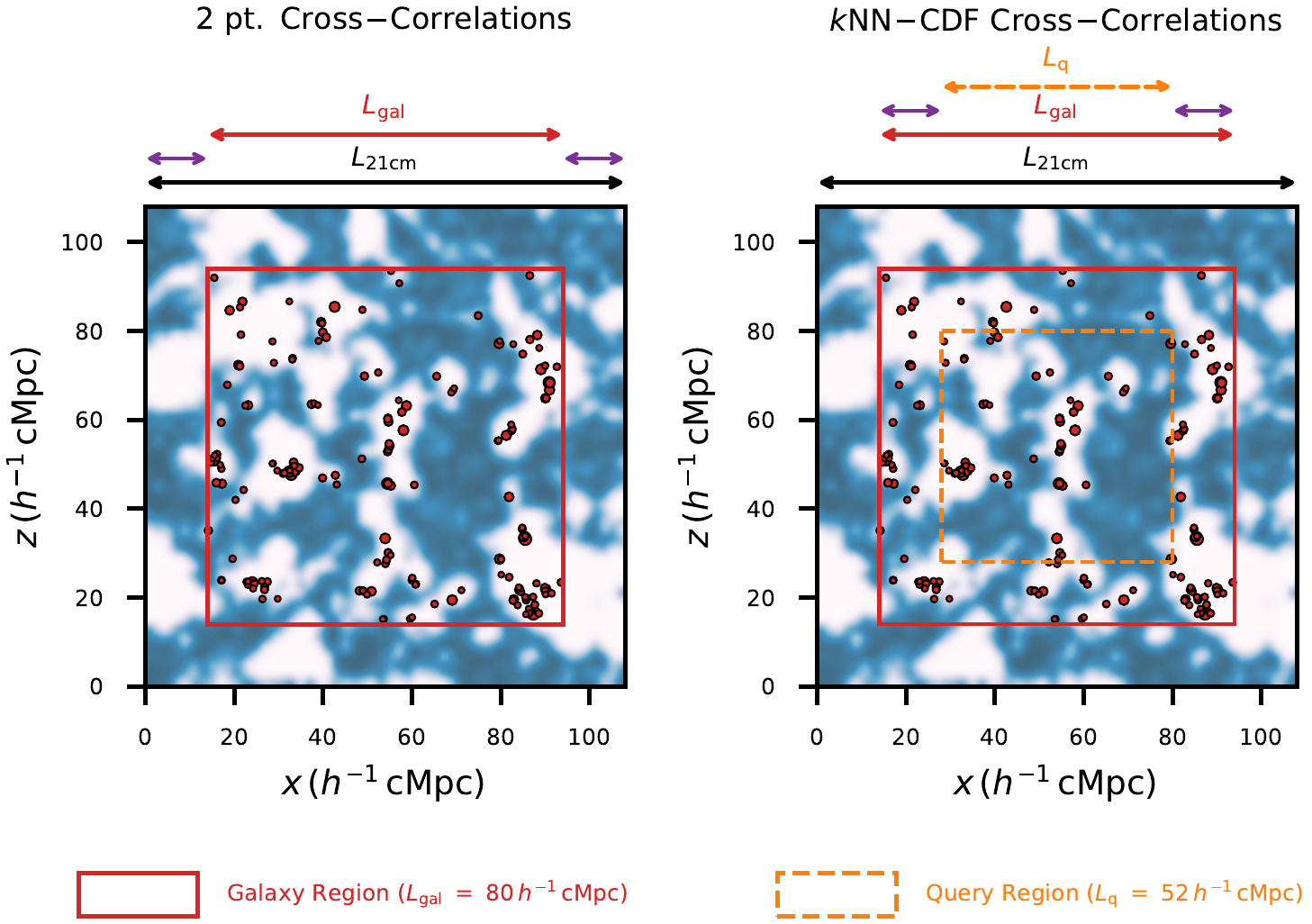}
    \caption{Geometric setup used in the 21\,cm--galaxy cross-correlation analyses. 
    The left panel shows the setup for the two-point cross-correlation analysis, in which the galaxy survey occupies a cubic volume of side length $L_{\rm gal}=80\,\hcMpc$, while the 21\,cm field is defined on a larger cube of side length $L_{\rm 21cm}=108\,\hcMpc$, maintaining a buffer length of $14\,\hcMpc$ (shown using violet arrows)  on all sides to mitigate edge effects. 
    The right panel shows the corresponding setup for the case of the $k$-nearest-neighbour distributions ($k$NN CDF), where query points are uniformly distributed within an inner cubic sub-volume of side length $L_q=52\,\hcMpc$, centered on the galaxy survey region and separated from its boundaries by the same buffer length.
    }
    \label{fig:geometric_setup_crosscor}
\end{figure}

In such a setup, it is therefore necessary for the 21\,cm field to extend beyond the galaxy survey boundaries by a buffer length on all sides. This buffer length is set by the largest scale $r_{\rm max}$ of interest, so that spherical shells can be fully constructed even for galaxies located near the boundaries of the survey region. In this work, as we measure the cross-correlation out to a maximum separation of $r_{\rm max}=12\,\hcMpc$, we use a buffer length of $14\,\hcMpc$ on each side of the galaxy survey volume. As a result, the 21\,cm field defined on a cubic volume of side length $108\,\hcMpc$, centered on the galaxy survey region, is used for the analysis. A schematic illustration of this geometric setup is shown in the left panel of \fig{fig:geometric_setup_crosscor}.

\subsection{Nearest Neighbour Cumulative Distribution Functions}
\label{subsec:kNN_cdf}

The $k$ nearest-neighbour cumulative distribution function ($k\mathrm{NN}$ CDF) formalism \cite{Banerjee2021_kNNIntro} offers an alternative approach to characterizing clustering in both discrete and continuous datasets, as well as for quantifying the cross-correlations between them.

For a dataset of $N$ discrete tracers distributed over a volume $V$ with mean number density $\bar n = N/V$, the $k\mathrm{NN}$ CDF at a length scale $r$ is defined as the volume-averaged probability, $\mathcal{P}_{\geq k}(r)$, that a randomly placed sphere of radius $r$ contains \emph{at least} $k$ tracers:
\begin{equation}
\label{knn_cdf}
\mathrm{CDF}_{k\mathrm{NN}}(r) = \mathcal{P}_{\geq k}(r)
\end{equation}
Unlike traditional two-point statistics, the $k\mathrm{NN}$ CDFs, by construction, is sensitive to all $N$-point correlations of the discrete tracers \cite{Banerjee2021_kNNIntro}, 

This formalism was subsequently extended to continuous fields by Banerjee \& Abel (2023) \cite{Banerjee2023_tracerfieldCross}. In the limit where a continuous field is represented by an infinitely dense sampling of discrete proxy points, with the mean number density $\bar{n} \rightarrow \infty$,  the volume-averaged probability, $\mathcal{P}_{\geq k}(r)$, of finding atleast $k$ data points in spheres of radius $r$ gets mapped to the probability, $\mathcal{P}_{> \rho^*_r}(r)$, that the underlying continuous field smoothed on a scale $r$ exceeds a threshold value $\rho^{*}_r$. Mathematically, we have
\begin{align}
    \mathcal{P}_{\geq k} (r) ~\longrightarrow~ \mathcal{P}_{> \rho^*_r}(r)~\propto~ \int_{\rho_r^{*}}^{\infty}
\Phi(\rho_r)\,{\rm d}\rho_r
=
1 - \mathrm{CDF}\!\left(\rho_r^{*}\right) .
\end{align}
where $\rho_r$ represents the continuous field smoothed on a radius $r$ using a spherical top-hat function and $\Phi(\rho_r)$ is the probability distribution function of the possible values of the smoothed field. 

In other words, for a given value of $r$, one can establish a direct correspondence between the discrete and continuous descriptions by associating the earlier nearest-neighbour index $k$ with some threshold value $\rho_r^*$ of the smoothed continuous field, defined through 
\begin{equation}
\dfrac{k}{ \bar{n} \left( \frac{4}{3}\pi r^3 \right) } = \rho_r^* 
\end{equation}
At fixed $r$, increasing the value of $k$ in the discrete case is thus equivalent to increasing the threshold $\rho_r^{*}$ in the continuous-field description. 

In this sense, the continuum analogue of the $k\mathrm{NN}$ measurements at scale $r$ essentially corresponds to evaluating the \emph{complementary} CDF of the \emph{smoothed} continuous field, from which the proxy data points were drawn, at the corresponding threshold value. As in the discrete 
$k\mathrm{NN}$ case,  the CDF of the smoothed continuous field also encodes contributions from correlations of all orders.

Building on this, Banerjee \& Abel (2023) \cite{Banerjee2023_tracerfieldCross} further showed that this formalism can be naturally extended to probe spatial cross-correlations between a discrete tracer population -- modelled as a local Poisson sampling of an underlying field $\tilde{\rho}(\boldsymbol{x})$ \footnote{Note that the field
$g(\boldsymbol{x})$ defined  earlier in
\eqn{eq:galaxy_field} is an overdensity field constructed from a Poisson-sampled tracer realization of this underlying continuous field $\tilde{\rho}(\boldsymbol{x})$.} with mean number density $\bar n$ -- and a continuous field $\rho(\boldsymbol{x})$. Within the nearest-neighbour framework, the cross-clustering between these two datasets on a length scale $r$ is measured by computing the joint probability of finding \emph{at least} $k$ discrete data points within a sphere of radius $r$ \emph{and} the density of the continuous field, smoothed over the same radius using a spherical top-hat function, exceeding a threshold $\rho_r^*$. For a given value of $k$ and a chosen threshold $\rho_r^*$, this joint probability, $\mathcal{P}_{\geq k, >\rho_r^*}$, can formally be expressed as:
\begin{equation}
\label{eq:joint_CDF_def}
\mathcal{P}_{\geq k, > \rho_r^*} = \mathcal{P}_{>\rho_r^*} - \sum_{j<k}\mathcal{P}_{j, >\rho_r^*},
\end{equation}
where $\mathcal{P}_{>\rho_r^*}$ denotes the marginal probability that the
continuous field, smoothed over a sphere of radius $r$, exceeds the threshold
$\rho_r^{*}$, irrespective of the number of discrete tracer points it contains, and the second term accounts for all those configurations in which the continuous field, smoothed on scale $r$, exceeds the threshold
$\rho_r^{*}$ but contains fewer than $k$ discrete tracer points within a sphere of radius $r$.

As shown in Banerjee \& Abel (2023) \cite{Banerjee2023_tracerfieldCross}, the quantity, $\mathcal{P}_{k,>\rho_r^*}$, which enters the summation in \eqn{eq:joint_CDF_def}  and corresponds to the joint probability of finding \emph{exactly} $k$ tracer points \emph{and} the
smoothed continuous field to cross the threshold $\rho^*_r$ within spheres of radius $r$, can be expressed explicitly as
\begin{equation}
\label{joint_prob_exact_k_def}
\mathcal{P}_{k, >\rho_r^*} =  \int_{\rho_r^*}^{\infty} \int_{0}^{\infty} \frac{\left[\lambda(\tilde \rho_r)\right]^k}{k!} e^{-\lambda(\tilde \rho_r)} \Phi(\tilde \rho_r, \rho_r)\, \mathrm{d}\tilde \rho_r \, \mathrm{d}\rho_r,   
\end{equation}
where, 
\begin{equation}
\label{eq:lambda}
\lambda(\tilde \rho_r) = \bar n \, \Bigg (\frac{4 \pi r^3}{3}\Bigg) \Bigg(\frac{\tilde \rho_r}{\langle \tilde \rho_r\rangle}\Bigg),
\end{equation}
and $\Phi(\tilde \rho_r,  \rho_r)$ represents the full joint probability distribution of the two fields $\tilde{\rho}$ and $\rho$ when smoothed on a scale $r$, and contains the complete information about the correlations between the fluctuations of the two fields.

Thus, it immediately follows that the joint probability, $\mathcal{P}_{\geq k, >\rho_r^*}$, defined in \eqn{eq:joint_CDF_def}, is also formally sensitive to correlations present in the fluctuations of the discrete tracers and the continuous field at all orders \cite{Banerjee2023_tracerfieldCross}.

If these two fields under consideration are statistically independent (i.e., completely uncorrelated), their joint probability distribution
$\Phi(\tilde \rho_r,  \rho_r) $ factorizes into the product of the individual probability distributions of the corresponding smoothed fields - 
\begin{equation}
\Phi(\tilde \rho_r,  \rho_r) \propto \Phi(\tilde \rho_r) \cdot\Phi( \rho_r) 
\end{equation}
In such a case, the joint probability of \eqn{eq:joint_CDF_def} simplifies to
\begin{equation}
\label{eq:joint_CDF_noCorr}
\mathcal{P}_{\geq k, > \rho_r^*} = \mathcal{P}_{\geq k} \times \mathcal{P}_{>\rho_r^*}~,
\end{equation}

This motivates the construction of a summary statistic that measures the \emph{excess} cross-correlation between a set of discrete tracers and a continuous field, independent of their auto-clustering. Thus, we define
\begin{equation}
\psi^{\mathrm{cross}}_{k\mathrm{NN}}(r)
\equiv
\frac{\mathcal{P}_{\ge k,>\rho_r^{*}}}
{\mathcal{P}_{\ge k}\,\times\mathcal{P}_{>\rho_r^{*}}},
\label{eq:psi_kNN_def}
\end{equation}
as the measure of the cross-correlation, noting that this statistic is also sensitive to all higher-order cross-correlations between the two datasets. Deviations of $\psi_{k\rm{NN}}$ from unity will indicate the  existence of spatial
cross-correlations, with values of $\psi_{k\rm{NN}} > 1$ ($\psi_{k\rm{NN}} < 1$) implying that the set of discrete tracers is correlated (anti-correlated) with the continuous field, while $\psi_{k\rm{NN}}  = 1$ indicates the \emph{complete} absence of \emph{any} spatial cross-correlation.

It is straightforward to understand that $k$NN statistics would be sensitive to the mean number density ($\bar{n}$) of tracers, since they are constructed from distances to the $k^{\rm th}$ nearest data points, which is directly set by the number of data points populating the given volume. As $\bar{n}$ increases, the typical $k\mathrm{NN}$ distances decrease, causing the cumulative distribution function ($k$NN CDF) to rise more rapidly and saturate to unity at smaller radii, while lower number densities lead to a more slowly saturating $k$NN CDF that extends to larger scales. Consequently, variations in the number density of the discrete tracer points alone can alter the shape of the $k$NN CDFs, even in the absence of changes in the underlying clustering. To robustly probe cross-correlations over the range of spatial scales: $4~\hcMpc \leq r \leq 12~\hcMpc$, we therefore adopted a sub-sampling strategy that corresponds to some fixed value of $\bar{n}$. Specifically, the full galaxy catalog, containing $N_{\rm gal}=7546$ objects, was partitioned into eight disjoint subsets of $N_{\rm gal,set}= 943$ galaxies each, constructed via random sampling without replacement from the parent catalog, with two unsampled galaxies excluded from the analysis. The $k$NN statistics are first computed independently for each subset, and the final measurement is then obtained by averaging over all eight subsets.

To numerically compute the cross-correlation signal between \emph{each such subset} of discrete
tracer points and the continuous field within the $k$NN formalism, we follow the procedure described in Banerjee \& Abel (2023) \cite{Banerjee2023_tracerfieldCross} with some modification to account for the non-periodic boundary conditions in our setup. The numerical implementation proceeds as follows:

\begin{enumerate}
    \item We generate $N_q^3$ query points, uniformly distributed within a cubic
    sub-volume of side length $L_q$ that is fully enclosed within the volume
    ($L_{\rm gal}^3$) occupied by the discrete tracer points
    (see right panel of \fig{fig:geometric_setup_crosscor}).
    
    More specifically, we place query points exclusively within an inner cubical volume, maintaining a buffer region of length $14~\hcMpc$ on all sides that separates them from the edges of the discrete tracer volume.  
    This ensures that tracer points are available to each query point out to the maximum length scale probed in this study (= $12 \hcMpc$) without encountering the boundaries of the tracer volume, which would otherwise artificially truncate nearest-neighbour distances.
    
    In this work, we choose $N_q^3 = 100^3$, ensuring that the number of query points is significantly larger than the number of discrete tracers in each subset ($N_{\rm gal,set}= 943$). We have verified that our results are stable with respect to the number of query points by repeating the analysis with $N_q^3 = 200^3$.

    \item We construct a $k$-d tree from the positions of the discrete tracers and use it to compute the distance from each query point to its $k$ nearest neighbours.
    For a given $k$, these distances are sorted to form the empirical cumulative
    distribution function, $\mathrm{CDF}_{k\mathrm{NN}}(r)$, which converges to
    $\mathcal{P}_{\ge k}(r)$ in the limit of large $N_q^3$.

    \item Next, we smooth the continuous field, itself defined on a grid of $N_{\rm g}^3$ points,
     with a spherical top-hat filter of radius $r$. The resulting smoothed
    field is then interpolated to the locations of the query points
    generated in step~1.

    \item We then select a threshold value $\rho_r^\ast$ for the smoothed field and compute the fraction of query points for which the interpolated smoothed field, $\rho_r$, exceeds this threshold.
    This fraction provides an empirical estimate of $\mathcal{P}_{>\rho_r^\ast}$.

    In this study, we follow the ``constant percentile'' approach of
    Banerjee \& Abel (2023) \cite{Banerjee2023_tracerfieldCross} and set $\rho_r^\ast$ to the value of the 75th percentile of $\rho_r$, denoted by $\rho_r^{75}$.  With this choice, $\mathcal{P}_{>\rho_r^{*}}$ is fixed to $0.25$, independent of the smoothing
scale $r$.
    
    \item  We compute the fraction of query points that \emph{simultaneously} satisfy the following two conditions:
    \begin{enumerate}
        \item the distance to the $k$-th nearest neighbour tracer point is less than or equal to $r$,
        \item the interpolated smoothed field at these query positions exceeds $\rho_r^\ast$.
    \end{enumerate}
    In the limit of large $N_q^3$, this fraction provides an estimate of
    $\mathcal{P}_{\ge k, > \rho_r^\ast}$.

    \item Finally, using these probabilities, we calculate the summary statistic
    $\psi^{\mathrm{cross}}_{k\mathrm{NN}}$, as defined in \eqn{eq:psi_kNN_def}.
    
    \item We repeat the procedure described in steps (3)--(6) for a range of smoothing scales $r$.

\end{enumerate}

In this work, we focus exclusively on the $k = 1$ case of the nearest-neighbour distribution, and defer a detailed investigation of using the higher orders to future work (however, see \app{appendix:addition_of_2NN_for_xcorr_detection} for an initial exploration in this direction).


\section{Prospects for Detecting Cross-Correlations Using Different Frameworks} 
\label{sec:detection_prospects}

In this section, we examine the prospects for detecting the cross-correlation signal between $\OIII$-emitting galaxies and the differential 21\,cm brightness temperature ($\delta T_{21}$) field, using the two frameworks described in \secn{sec:crosscorr_frameworks}.

\subsection{Analysis Methodology}
\label{subsec:method_detection_prospects}

We use mock observed catalogs of $\OIII$-emitting galaxies and 21\,cm datasets
simulated following the \textbf{fiducial} model described in
\secns{subsec:galaxy_model}{subsec:reion_21cm_model}. The input galaxy catalog is held fixed throughout the analysis, while the simulated $\delta T_{21}$ field is modified to represent different 21\,cm observational scenarios. We consider the following two cases:
\begin{itemize}
\item \textbf{Noise}: Here, we construct a \emph{noisy} 21\,cm data cube by adding
instrumental noise representative of SKA-Low observations (see
\secn{subsec:reion_21cm_model} for details) to the $\delta T_{21}$ field obtained from
\texttt{SCRIPT}, with no foreground filtering applied. This represents an
idealized scenario in which foreground contamination is assumed to be perfectly removed.

\item \textbf{Noise + Foreground Filtering}: In this case, the \emph{noisy} 21\,cm data cube obtained for the previous case is further processed to mitigate contamination from
foreground emission. We adopt a pessimistic foreground-avoidance strategy, as
described in \secn{subsec:reion_21cm_model}, by applying the wedge filter defined in \eqn{eq:wedge_equation}.
\end{itemize}

Together, these two cases bracket the range of effective data quality expected from realistic interferometric 21\,cm experiments, and allow us to assess how robustly the 21\,cm--galaxy cross-correlation can be measured in the presence of
instrumental noise and foreground contamination.

We adopt a hypothesis-testing approach to assess the presence of 21\,cm--galaxy cross-correlations, with the null hypothesis corresponding to the absence of any statistically significant spatial correlation between the galaxy distribution and fluctuations in the 21 cm signal. To accept or reject the null hypothesis, it is necessary to construct a control sample that satisfies its assumptions. For this purpose, we generate a catalog of randomly distributed galaxies within the same survey volume, matched to the number density of the mock galaxy catalog. Cross-correlating these randomized galaxy catalogs with 21\,cm noise cubes -- optionally, including foreground wedge filtering, depending on the case at hand -- provides an estimate of the cross-correlation expected if the null hypothesis is true. We generate $N_{\rm null}$ independent realizations of such tracer–field datasets, which together define the control sample used in the statistical inference.

Let $S(r)$ denote a chosen summary statistic -- either the two-point
cross-correlation function $\xi_{\rm cross}(r)$ or the $1\mathrm{NN}$ CDF excess cross-correlation statistic $\psi^{\mathrm{cross}}_{1\mathrm{NN}}(r)$ -- evaluated at a discrete set of $m$ length scales $\{ r_1, r_2, \ldots, r_m \}$. We define a \emph{data vector} $\boldsymbol{\mathcal{D}}_{\rm S}$ from measurements of this statistic in the mock observed dataset as follows:
\begin{equation}
\boldsymbol{\mathcal{D}}_{\rm S}
\equiv
\left(
\mathcal{D}_{{\rm S},1},
\mathcal{D}_{{\rm S},2},
\ldots,
\mathcal{D}_{{\rm S},m}
\right),
\end{equation}
where $\mathcal{D}_{{\rm S},i}$ represents the value of the statistic $S$ evaluated at the length
scale $r_i$ and $m$ is the length of the data vector.

Similarly, for each control dataset, we construct a
corresponding \emph{null data vector}
\begin{equation}
\boldsymbol{\mathcal{N}}^{\alpha}_{\rm S}
\equiv
\left(
\mathcal{N}^{\alpha}_{{\rm S},1},
\mathcal{N}^{\alpha}_{{\rm S},2},
\ldots,
\mathcal{N}^{\alpha}_{{\rm S},m}
\right),
\end{equation}
where $\mathcal{N}^{\alpha}_{{\rm S},i} = S(r_i)$ and the index
$\alpha ~(= 1,2,\ldots,N_{\rm null})$ labels independent realizations drawn under the null hypothesis.

To quantify the statistical significance of the measured cross-correlation, we
compute a $\chi^2$ statistic for both the mock observed dataset and each 
realization of the control dataset:
\begin{align}
\chi^2_{\rm S} &=
\left( \boldsymbol{\mathcal{D}}_{\rm S}
- \langle \boldsymbol{\mathcal{N}}_{\rm S} \rangle \right)^{\mathrm{T}}
~\boldsymbol{\Sigma}^{-1}_{\rm S}~
\left( \boldsymbol{\mathcal{D}}_{\rm S}
- \langle \boldsymbol{\mathcal{N}}_{\rm S} \rangle \right), \\
\chi^2_{{\rm S},\alpha} &=
\left( \boldsymbol{\mathcal{N}}^{\alpha}_{\rm S}
- \langle \boldsymbol{\mathcal{N}}_{\rm S} \rangle \right)^{\mathrm{T}}
~\boldsymbol{\Sigma}^{-1}_{\rm S}~
\left( \boldsymbol{\mathcal{N}}^{\alpha}_{\rm S}
- \langle \boldsymbol{\mathcal{N}}_{\rm S} \rangle \right),
\label{eq:chisq_detection}
\end{align}
Here, $\langle \boldsymbol{\mathcal{N}}_{\rm S} \rangle$ denotes the ensemble mean of
the null data vectors, and $\boldsymbol{\Sigma}^{-1}_{\rm S}$ is the inverse covariance
matrix of the summary statistic $S$, estimated from the control sample as described
below.

Using the set of $N_{\rm null}$ null data vectors
$\{\boldsymbol{\mathcal{N}}^{\alpha}_{\rm S} \}$, we first compute the covariance matrix $\boldsymbol{\tilde{\Sigma}}_{\rm S}$, whose elements are given by
\begin{equation}
\tilde{\Sigma}_{\rm S}(i,j)
=
\Big\langle
\left( \mathcal{N}^{\alpha}_{{\rm S},i}
- \langle \mathcal{N}_{{\rm S},i} \rangle \right)
\left( \mathcal{N}^{\alpha}_{{\rm S},j}
- \langle \mathcal{N}_{{\rm S},j} \rangle \right)
\Big\rangle,
\label{eqn:cov_matrix}
\end{equation}
where the angular brackets denote an average over the $N_{\rm null}$ realizations. 
Since the covariance matrix is estimated from a finite number of realizations,
we apply the Hartlap correction \cite{Hartlap2007} to obtain an unbiased estimate of its inverse:
\begin{equation}
\boldsymbol{\Sigma}^{-1}_{\rm S}
=
\frac{N_{\rm null} - m - 2}{N_{\rm null} - 1}
~~
\boldsymbol{\tilde{\Sigma}}_{\rm S}^{-1},
\end{equation}
which is then used to compute the $\chi^2$ statistic via \eqn{eq:chisq_detection}.

Finally, the set $\chi^2_{{\rm S,null}} \equiv \{\chi^2_{{\rm S},\alpha}\}$ represents the
\emph{null distribution}, which characterizes the level of signal-to-noise
expected under the null hypothesis and serves as the reference against which
the value obtained from the mock observations, $\chi^2_{\rm S}$, is evaluated.
A large value of $\chi^2_{\rm S}$ relative to the null distribution indicates a higher statistical detection significance of the cross-clustering signal. It is worth mentioning that tests of the $k$NN statistics in our previous works have shown that the likelihood is well approximated by a multivariate Gaussian, provided one stays away from the tails \cite{Wang2022, Zhou2025_KNN, Chand2025}. Consequently, the use of a standard $\chi^2$ analysis is well justified.

\begin{figure}
    \centering
    \includegraphics[width=\linewidth]{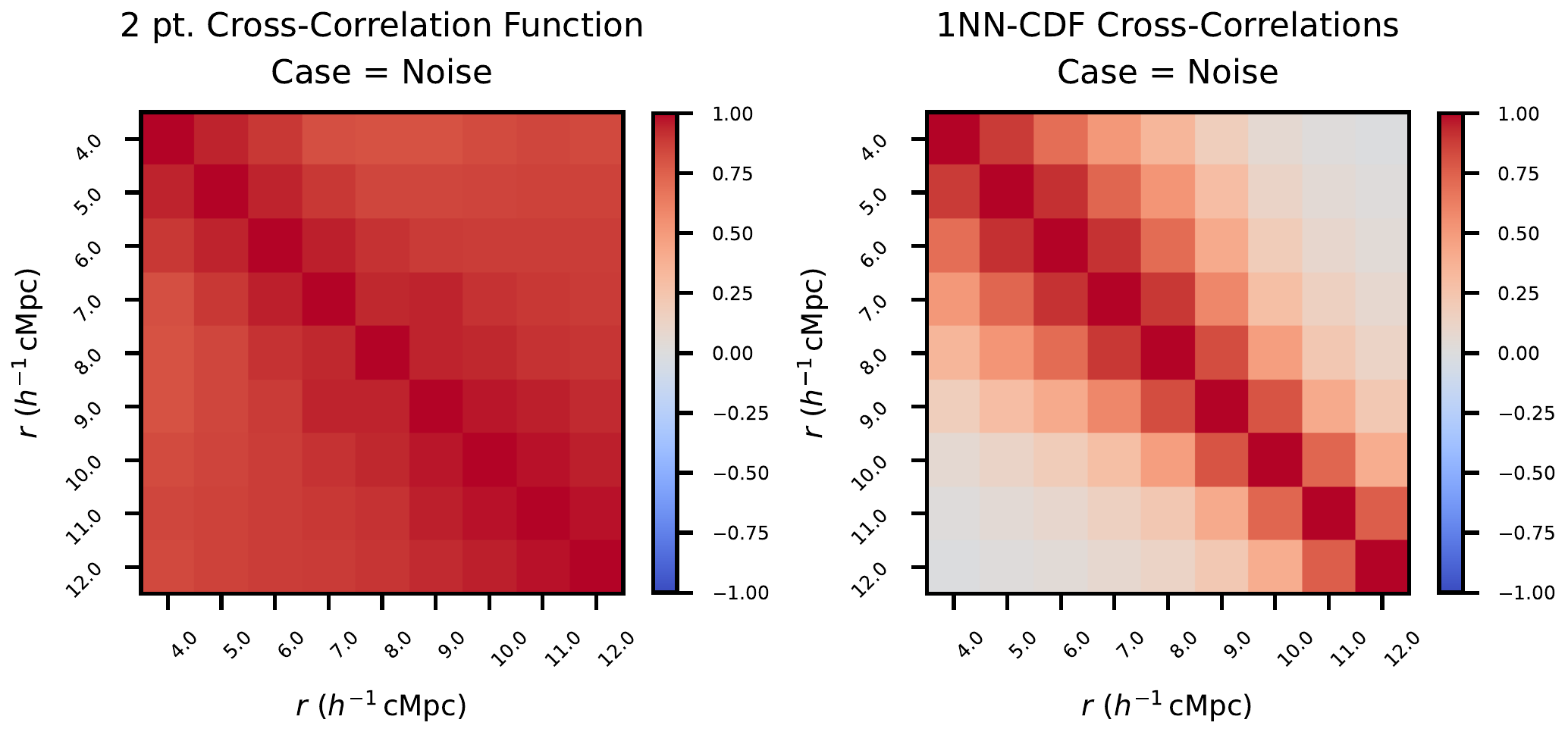}

    \vspace{0.35cm}

    \includegraphics[width=\linewidth]{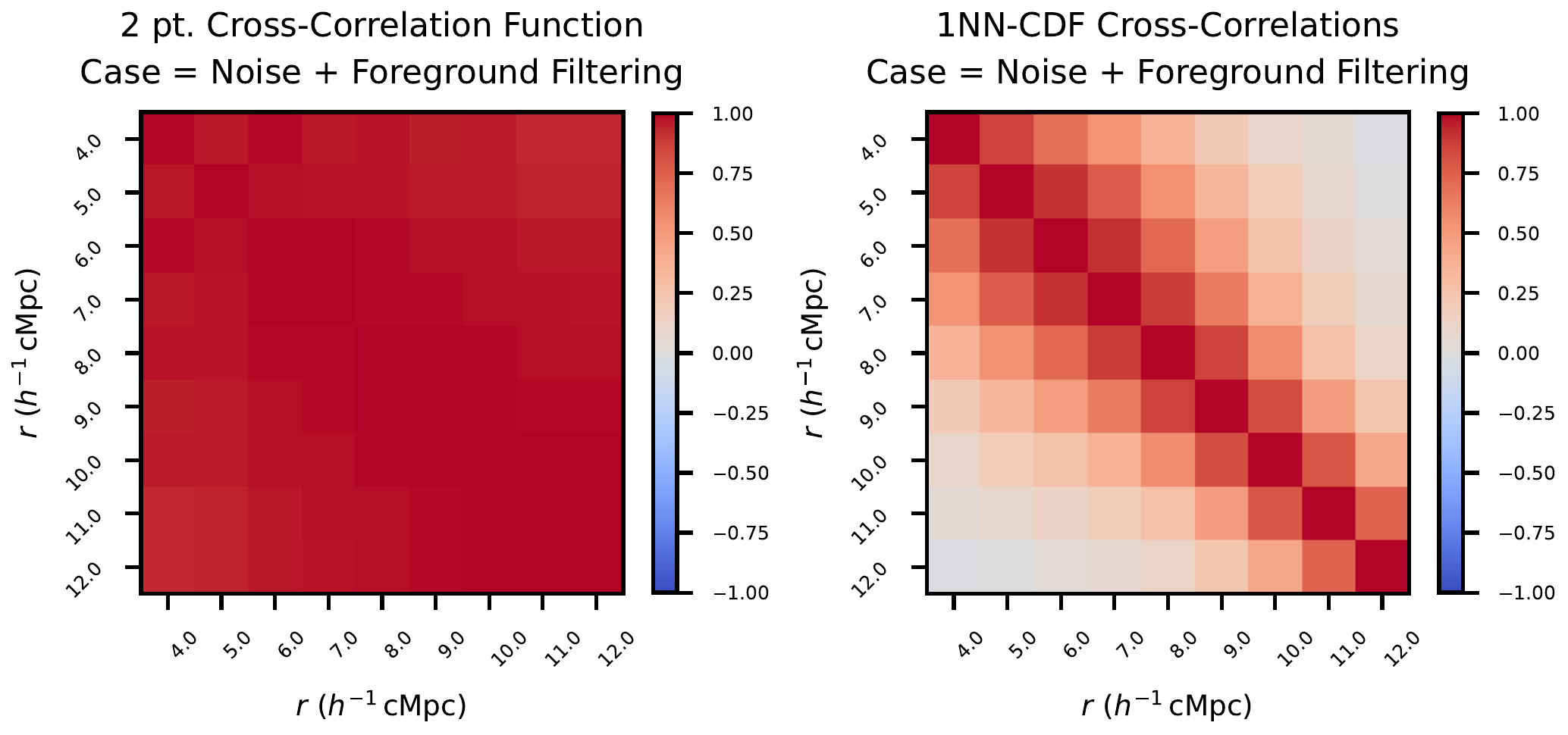}

    \caption{
    The correlation matrices corresponding to the different observational scenarios discussed in \secn{subsec:method_detection_prospects}. The \emph{top row} presents the results for the \textbf{Noise} scenario, while the \emph{bottom row} corresponds to the \textbf{Noise + Foreground Filtering} scenario. These matrices have been calculated from an ensemble of $N_{\rm null} = 1000$ realizations of the null hypothesis.
    }
    \label{fig:corr_matrices}
\end{figure}

\fig{fig:corr_matrices} shows the structure of correlation matrices for both the
two-point cross-correlation function and the 1NN CDF excess cross-correlation statistics, computed from
1000 realizations of the null hypothesis. The top panels correspond to the
\textbf{Noise} case, while the bottom panels show the \textbf{Noise + Foreground Filtering} case. We find that the correlation matrix of the two-point 21\,cm--galaxy cross-correlation function exhibits significant, uniformly positive off-diagonal terms.  However, compared to the two-point case, the covariance of the 1NN cross-correlation statistic is more strongly localized around the diagonal, with correlations primarily confined to neighbouring radial bins.

\subsection{Results}
\label{subsec:results_detection_prospects}


\begin{figure}[htbp]
    \centering
    \begin{subfigure}{\columnwidth}
        \centering
        \includegraphics[width=\columnwidth]{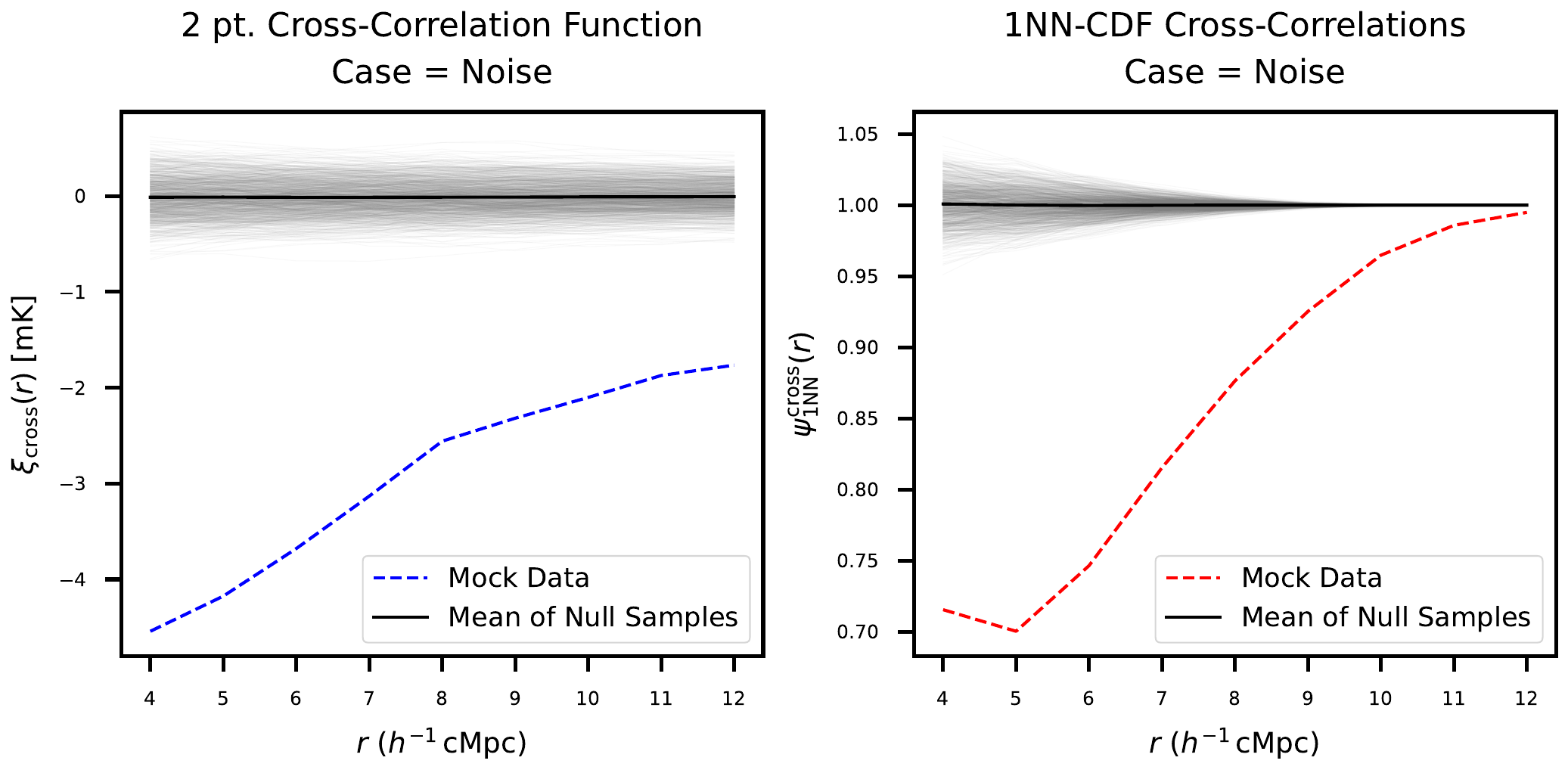}
    \end{subfigure}

    \vspace{0.5cm}

    \begin{subfigure}{\columnwidth}
        \centering
        \includegraphics[width=\columnwidth]{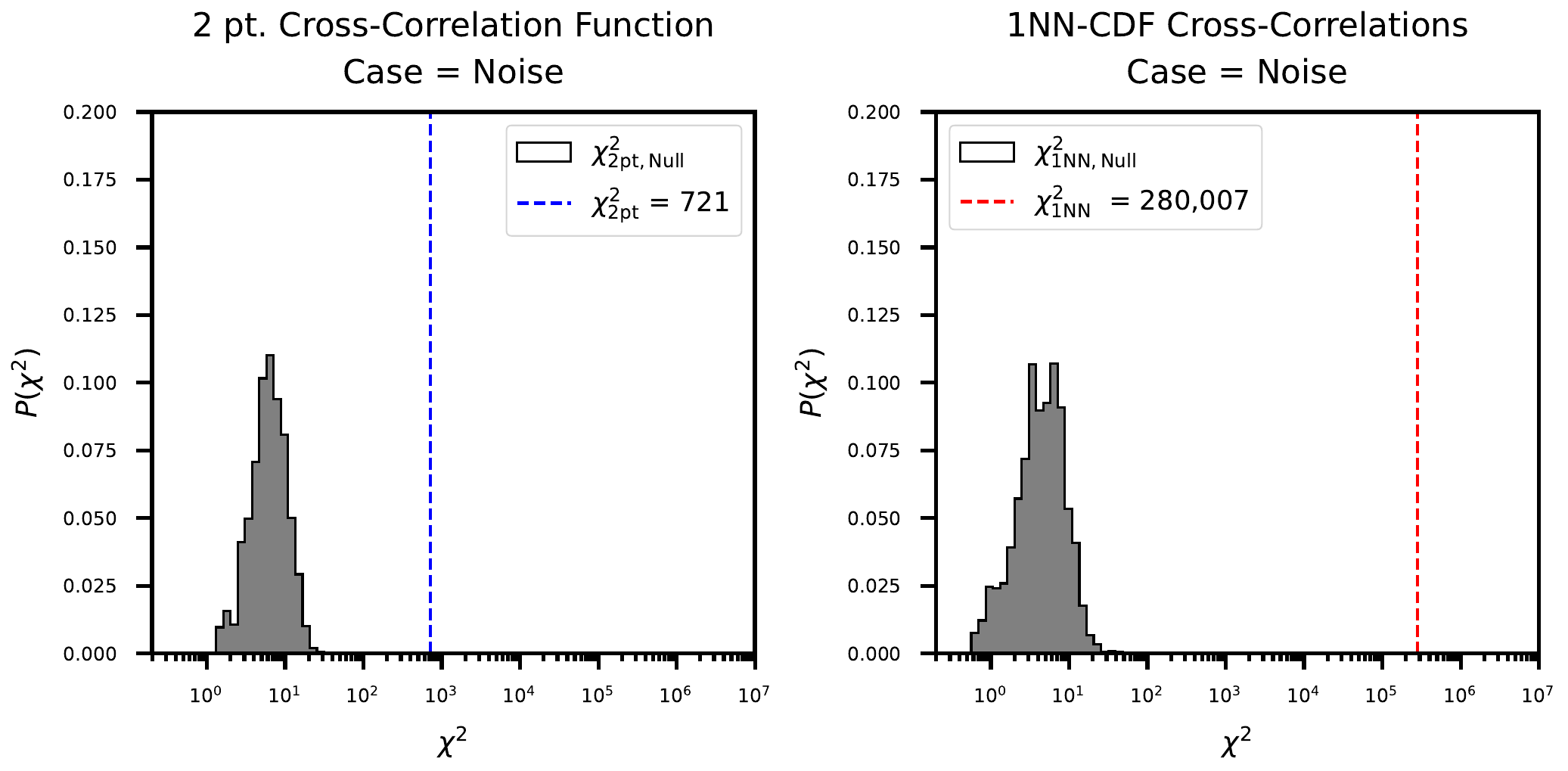}
    \end{subfigure}

    \caption{
    Prospects for detecting 21\,cm--galaxy cross-correlations in the \textbf{Noise} scenario (see \secn{sec:detection_prospects} for details).\\
    \textbf{\textit{Top row:}} The cross-correlation summary statistic measured from the mock dataset (colored dashed curves) and the $N_{\rm null} =$ 1000 control datasets (gray solid curves), shown for the two-point cross-correlation function (left panel) and the first nearest-neighbour (1NN) CDF cross-correlation statistic (right panel). \textbf{\textit{Bottom row:}} Distributions of the $\chi^2$ statistic corresponding to the cross-correlation measurement from the mock dataset (vertical dashed lines) and the 1000 control datasets (gray histograms), computed using the two-point (left panel) and the 1NN CDF cross-correlation statistics (right panel).
    }
    \label{fig:detect_prospect_noise}
\end{figure}

We begin by discussing the results for the \textbf{Noise} case, which are shown in \fig{fig:detect_prospect_noise}.

As shown in the panels on the top row of \fig{fig:detect_prospect_noise}, both summary statistics reveal an anti-correlation between galaxies and the 21\,cm field in the simulated mock observations, with $\xi_{\mathrm{cross}} < 0$ and $\psi^{\mathrm{cross}}_{1\mathrm{NN}} < 1$ over the separations considered here. This anti-correlation arises because galaxies preferentially reside within ionized regions during reionization, in which the local 21\,cm brightness temperature is suppressed relative to the mean. On scales larger than the typical sizes of ionized regions, the 21\,cm fluctuations become effectively uncorrelated with individual galaxies, causing both statistics to asymptotically approach their respective null signal as one moves to sufficiently large separations.

We remind the reader that the amplitude of the two summary statistics used in this work conveys fundamentally different information. The two-point 21 cm–galaxy cross-correlation function, $\xi_{\mathrm{cross}}(r)$, has a direct physical interpretation: its amplitude can be interpreted as the \emph{absolute} excess (or deficit) of the 21 cm brightness temperature measured at a separation $r$ from galaxy locations relative to the mean field evaluated around random locations in the volume. In contrast, the 1NN CDF cross-correlation statistic $\psi^{\mathrm{cross}}_{1\mathrm{NN}}$ does not measure an excess signal in the same sense. Rather, the amplitude of $\psi^{\mathrm{cross}}_{1\mathrm{NN}}$, which is defined as the ratio of joint to marginal nearest-neighbour probabilities (see \eqn{eq:psi_kNN_def}), simply represents a measure of the significance of the cross-correlation. Crucially, it should also be kept in mind that obtaining $\psi^{\mathrm{cross}}_{1\mathrm{NN}}=1$ indicates a \emph{complete} statistical independence between the two datasets, whereas  
a vanishing two-point cross-correlation, 
$\xi_{\mathrm{cross}}=0$, rules out the presence of correlations only at the two-point (linear) level and may still allow for higher-order statistical correlations that are not captured by this statistic. 

The statistical significance associated with the detection of 21\,cm--galaxy cross-correlations is further quantified in the bottom panels of \fig{fig:detect_prospect_noise}, where the $\chi^2$ distributions computed from control samples consistent with the null hypothesis of no cross-correlation are also shown (gray histograms) for both summary statistics. In each case, the $\chi^2$ value measured from the mock dataset (vertical dashed line) lies far away in the tail of the corresponding null distribution, indicating a strong inconsistency with the null hypothesis. Notably, the 1NN CDF statistic refutes the null hypothesis at substantially higher significance ($\chi^2_{\mathrm{1NN}} \approx 2.8 \times 10^{5}$) than the conventional two-point statistic ($\chi^2_{\mathrm{2pt}} = 721$), indicating that the 1NN CDF cross-correlation framework is markedly more robust to the underlying anti-correlation, even in the presence of instrumental noise. This higher detection significance in the case of the 1NN statistic essentially reflects its ability to capture non-Gaussian, higher-order cross-correlations that are not captured by the two-point cross-correlation function.


\begin{figure}[htbp]
    \centering
    \begin{subfigure}{\columnwidth}
        \centering
        \includegraphics[width=\columnwidth]{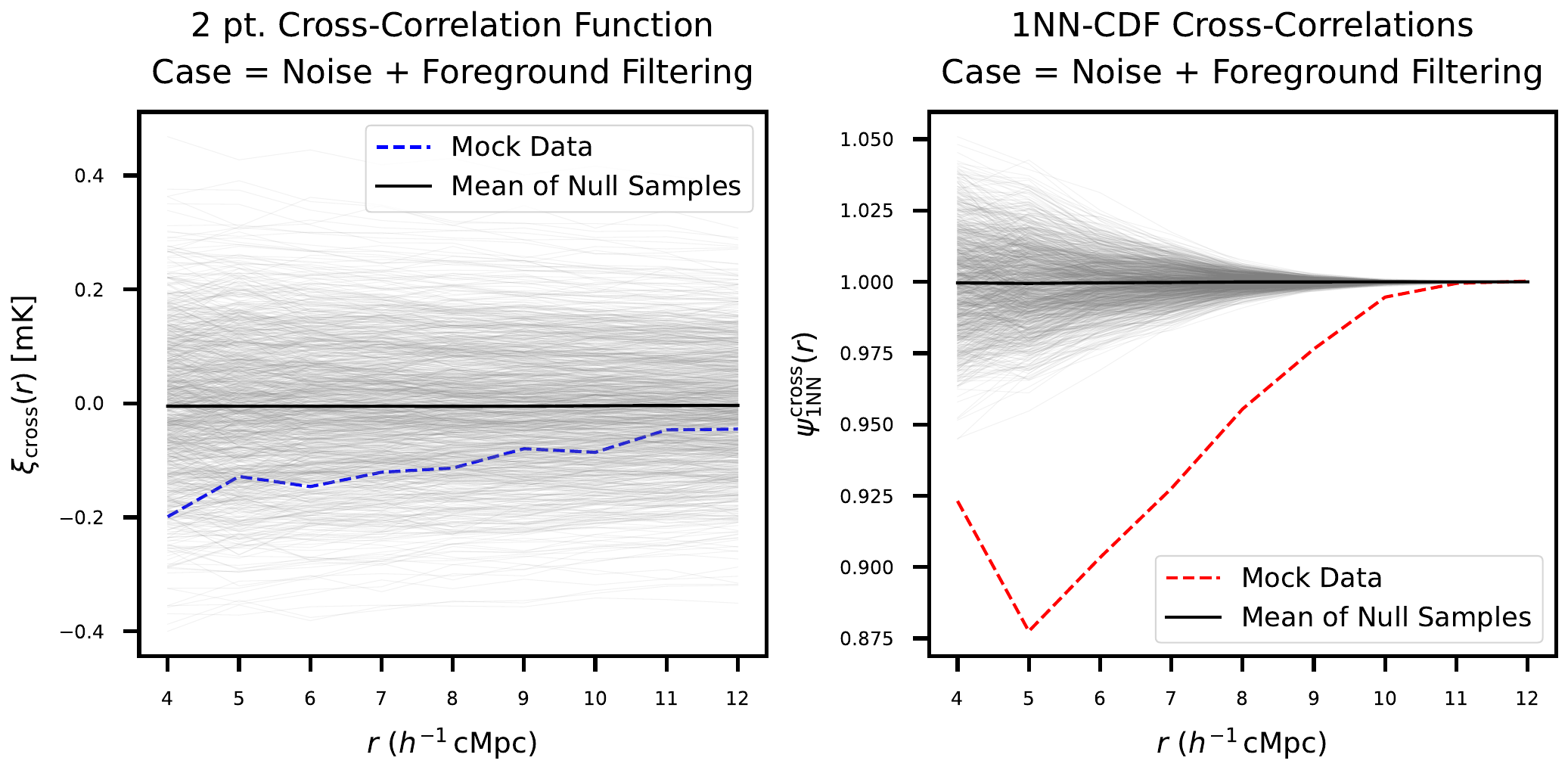}
    \end{subfigure}

    \vspace{0.5cm}

    \begin{subfigure}{\columnwidth}
        \centering
        \includegraphics[width=\columnwidth]{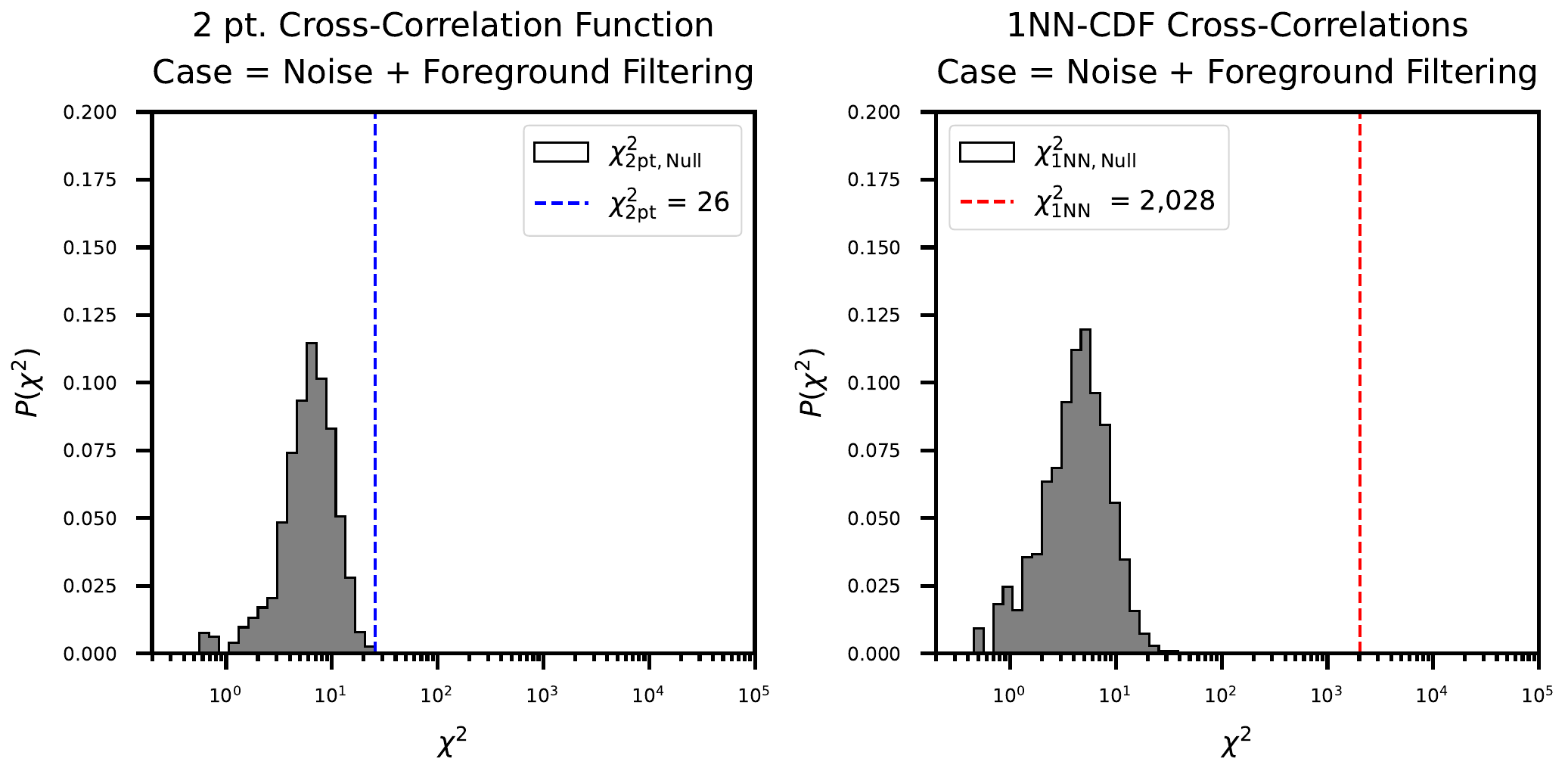}
    \end{subfigure}

    \caption{Same as \fig{fig:detect_prospect_noise}, but for the \textbf{Noise + Foreground Filtering} scenario.}
    \label{fig:detect_prospect_noise_plus_wedge}
\end{figure}

When foreground filtering is applied, the behavior of both summary statistics changes appreciably. As shown in the top row of \fig{fig:detect_prospect_noise_plus_wedge}, the amplitudes of both statistics are substantially reduced relative to the noise-only case. In particular, the two-point cross-correlation function measured from the mock dataset becomes largely indistinguishable from that of the control sample across all separations. This loss of sensitivity arises because the two-point statistic measure is dominated by large-scale modes at low $k_\parallel$, which are preferentially removed from the 21\,cm field by our foreground avoidance strategy\footnote{Formally, the two-point cross-correlation function can be expressed as
\begin{equation}
\xi_{\rm cross}(r)
=
\int_{0}^{\infty} \frac{k_\perp\, dk_\perp}{(2\pi)^2}
\int_{-\infty}^{+\infty} dk_\parallel \,
P_{\rm cross}(k_\perp,k_\parallel)\,
J_0(k_\perp r_\perp)\,
e^{i k_\parallel r_\parallel}.
\end{equation}
where $r = \sqrt{r_\perp^2 + r_\parallel^2}$, $P_{\rm cross}(k)$ is the cross-power spectrum and $J_0(x)=
\frac{1}{2\pi}
\int_0^{2\pi} d\phi \;
e^{i x \cos\phi}$ is the cylindrical Bessel function of order zero. The kernel $J_0(k_\perp r_\perp)\,e^{i k_\parallel r_\parallel}$ preferentially weights long-wavelength modes, particularly at low $k_\parallel$, where the 21\,cm--galaxy cross power is also expected to be concentrated. Removing this low-$k_\parallel$ region across a wide range of $k_\perp$, via foreground wedge avoidance, therefore leads to a significant suppression of the amplitude of $\xi_{\rm cross}(r).$}. Consequently, the underlying two-point cross-correlation signal that was present in the \textbf{Noise} case is now almost entirely washed out. In contrast, the 1NN CDF cross-correlation signal remains distinguishable from the null hypothesis even after foreground filtering, as seen in the top-right panel of \fig{fig:detect_prospect_noise_plus_wedge}. This behavior, also seen in Chand et al. (2025) \cite{Chand2025}, likely arises because foreground filtering does not completely erase higher-order correlations, to which the $k$NN statistic is inherently sensitive \cite{Banerjee2021_kNNIntro}. These trends can be understood more quantitatively by looking at the $\chi^2$ distributions shown in the bottom row of \fig{fig:detect_prospect_noise_plus_wedge}. In the \textbf{Noise + Foreground Filtering} case, the two-point statistic yields a modest value of $\chi^2_{\mathrm{2pt}} = 26$, indicating only marginal evidence for a cross-correlation. Whereas, the 1NN CDF statistic produces a substantially larger value, $\chi^2_{\mathrm{1NN}} \approx  2 \times 10^{3}$, placing it deeper into the tail of the null distribution and implying a clear detection of cross-correlations from the datasets.

These results demonstrate that nearest-neighbour distributions offer a more robust probe of 21\,cm--galaxy correlations under realistic observational setups. The 1NN CDF statistics, by virtue of their sensitivity to higher-order non-Gaussian information, retain significant detection power even when aggressive foreground mitigation is applied, making it a compelling alternative to conventional two-point statistics for cross-correlation studies during the EoR. While our analysis in this work is restricted to the case $k=1$, we further examine the impact of incorporating higher-order ($k>1$) nearest-neighbour statistics in \app{appendix:addition_of_2NN_for_xcorr_detection}.


\section{Distinguishing Reionization Scenarios using Cross-Correlations }
\label{sec:diff_reion_models}

In this section, we investigate the prospects of using the 21\,cm--galaxy cross-correlation frameworks described in \secn{sec:crosscorr_frameworks} to distinguish
between different reionization morphologies, and in particular, test whether the $k\mathrm{NN}$ CDF formalism provides improved discriminatory power over a standard two-point cross-correlation analysis.

\begin{figure}
\centering
\includegraphics[width=\columnwidth]{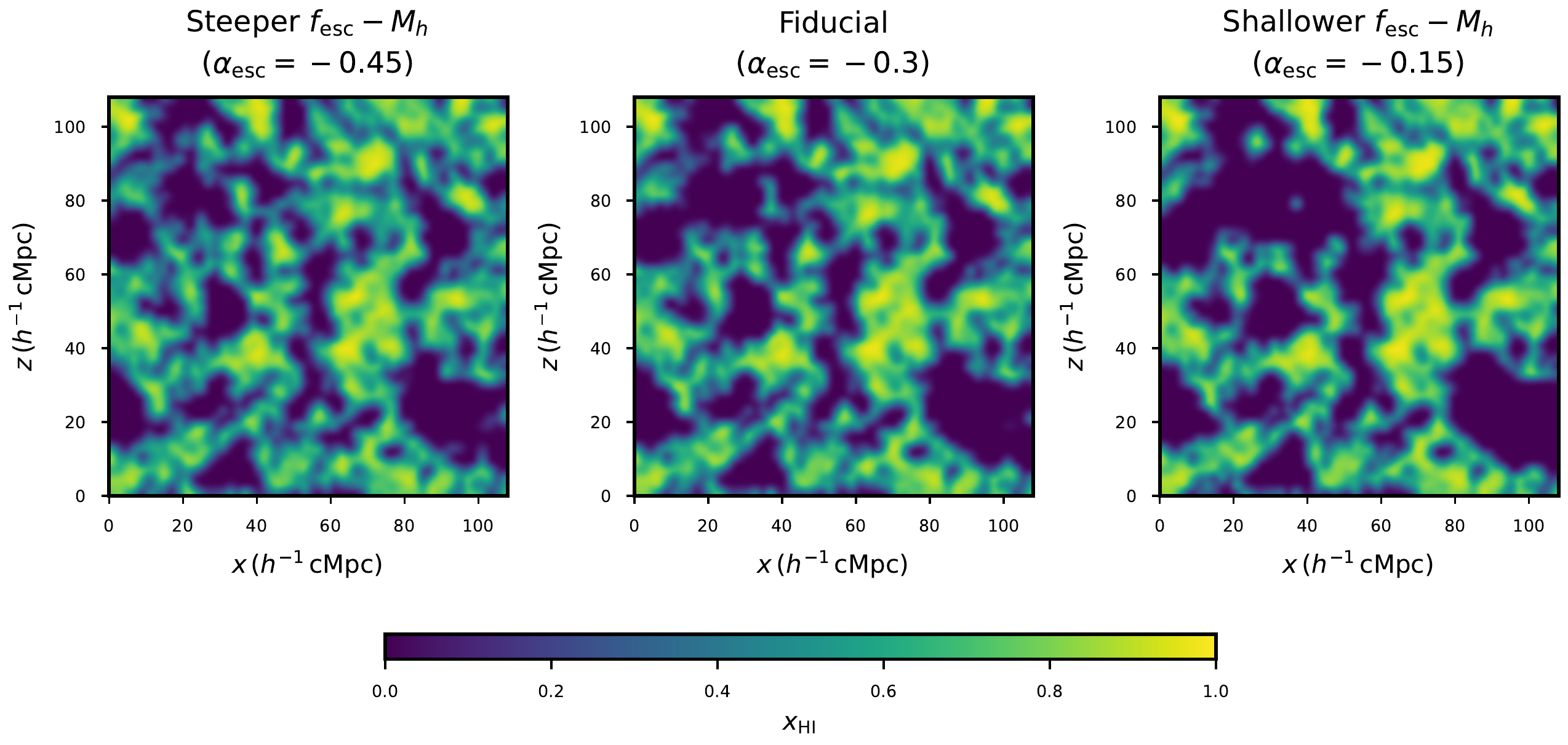}
\caption{Two-dimensional slices of the neutral hydrogen fraction field ($x_{\mathrm{HI},i} \equiv 1-x_{\mathrm{HII},i}$) at $z=7$ for three different reionization models with distinct ionizing escape fraction prescriptions: \textbf{steeper–$\bm{f_{\rm esc}}$} ($\alpha_{\mathrm{esc}}=-0.45$), \textbf{fiducial} ($\alpha_{\mathrm{esc}}=-0.3$), and \textbf{shallower–$\bm{f_{\rm esc}}$} ($\alpha_{\mathrm{esc}}=-0.15$). Each panel displays the same slice of thickness $2\,\hcMpc$ from the $108^3\,\cubedhcMpc$ sub-volume used in the cross-correlation analysis. Although all three models are normalized to yield the same mass-weighted, globally averaged neutral fraction $Q^{M}_{\mathrm{HI}}$ over the full simulation volume (see \secn{sec:diff_reion_models} for details), the morphology of ionized regions is noticeably different between the models.}
\label{fig:reion_model_comparision}
\end{figure}

To simulate different reionization scenarios with distinct ionization topology, we vary the prescription for assigning ionizing escape fraction $f_{\rm esc}(M_h)$ to halos in the model. The resulting 21\,cm fields then provide an ideal testbed for assessing the sensitivity of the two-point and $k\mathrm{NN}$ CDF cross-correlation frameworks to the astrophysical processes governing reionization. Our \textbf{fiducial} model assumes that the ionizing escape fraction follows a power-law dependence on halo mass with slope $\alpha_{\mathrm{esc}}=-0.3$.  In addition to this baseline case, we consider two alternative reionization scenarios:
\begin{itemize}
    \item \textbf{steeper-$\bm{f_{\rm esc}}$}: a scenario in which the ionizing escape fraction $f_{\mathrm{esc}}$ increases more rapidly toward lower halo masses, with a steeper power-law slope, $\alpha_{\mathrm{esc}}=-0.45$, 

    \item \textbf{shallower--$\bm{f_{\rm esc}}$}: a scenario in which the ionizing escape fraction $f_{\mathrm{esc}}$ increases more weakly toward lower halo masses, with a shallower power-law slope, $\alpha_{\mathrm{esc}}=-0.15$.
\end{itemize}

For each choice of $\alpha_{\mathrm{esc}}$, we adjust the normalization $f_{\mathrm{esc},10}$ of the
$f_{\mathrm{esc}}(M_h)$ relation such that the mass-weighted mean neutral hydrogen fraction,
$Q^{M}_{\mathrm{HI}}$, averaged over the full simulation volume, is identical across all the three
models at $z=7$. Note that in our theoretical framework, the ionizing efficiency of halos scales as
$\zeta \propto f_{\mathrm{esc}}(M_h)\,f_\star(M_h)$. For halo masses below $10^{12}\,M_\odot$\footnote{At higher masses, the star-formation efficiency $f_\ast$ itself declines
rapidly, and such halos become exponentially rare at the redshifts of interest, rendering their
overall contribution to the ionizing photon budget during reionization negligible.}, this scaling
implies $\zeta \propto M_h^{\alpha_{\mathrm{esc}}+\alpha_{\ast,\mathrm{lo}}}$. Consequently, for fixed $Q^{M}_{\mathrm{HI}}$, variations in $\alpha_{\mathrm{esc}}$ do not modify the
\emph{total} ionizing photon budget, but instead redistribute it across halo mass. This manifests as differences in the size and connectivity of ionized regions among the models,
as illustrated in \fig{fig:reion_model_comparision} -- where the \textbf{steeper-$\bm{f_{\rm esc}}$} model produces smaller, more fragmented ionized regions, while the \textbf{shallower-$\bm{f_{\rm esc}}$} model yields slightly larger, more connected ionized bubbles.

Since the $\OIIIang$ luminosity in our theoretical model scales with $(1 - f_{\mathrm{esc}})$ (see \eqn{eq:loiii_basic}), variations in the ionizing escape fraction would otherwise alter the predicted galaxy $\OIII$ luminosities and, consequently, the disrete tracer population. To prevent this, we recalibrate the normalization factor $\CfacOIII$ independently for each model, ensuring that the resulting catalogs of $\OIII$ emitters at $z=7$ are identical across all reionization scenarios and reproduce the same $\OIIIang$ luminosity function. With this calibration applied, any differences in the cross-correlation signal can be attributed exclusively to variations in the ionization morphology.

To quantitatively evaluate the ability of each statistical framework to
distinguish between the fiducial model and a given alternate model $\mathcal{M}$, we compute a $\chi^2$ statistic defined as
\begin{align}
\chi^2_{\mathrm{2pt};\,(\mathrm{fid},\,\mathcal{M})} 
&= 
\big(\boldsymbol{\xi}^{\mathrm{cross}}_{\mathrm{fid}} -
\boldsymbol{\xi}^{\mathrm{cross}}_{\mathcal{M}} \big)^{\mathrm{T}}
\,\boldsymbol{\Sigma}^{-1}_{\mathrm{2pt}}\,
\big(\boldsymbol{\xi}^{\mathrm{cross}}_{\mathrm{fid}} -
\boldsymbol{\xi}^{\mathrm{cross}}_{\mathcal{M}} \big), \\[6pt]
\chi^2_{\mathrm{1NN};\,(\mathrm{fid},\,\mathcal{M})} 
&= 
\big(\boldsymbol{\psi}^{\mathrm{cross}}_{\mathrm{1NN,fid}} -
\boldsymbol{\psi}^{\mathrm{cross}}_{\mathrm{1NN},\mathcal{M}} \big)^{\mathrm{T}}
\,\boldsymbol{\Sigma}^{-1}_{\mathrm{1NN}}\,
\big(\boldsymbol{\psi}^{\mathrm{cross}}_{\mathrm{1NN,fid}} -
\boldsymbol{\psi}^{\mathrm{cross}}_{\mathrm{1NN},\mathcal{M}} \big),
\label{eq:chisq_differentiate_defn}
\end{align}
where $\boldsymbol{\xi}$ denotes the data vector computed for the two-point
cross-correlation function and
$\boldsymbol{\psi}_{\mathrm{1NN}}$ denotes the data vector for the excess
nearest-neighbour cross-correlation statistic. The label $\mathcal{M}$ denotes either the
\textbf{shallower-$\bm{f_{\rm esc}}$} model
($\mathcal{M}=\mathrm{shallow}$) or the
\textbf{steeper-$\bm{f_{\rm esc}}$} model
($\mathcal{M}=\mathrm{steep}$). The inverse covariance matrices
$\boldsymbol{\Sigma}^{-1}_{\mathrm{2pt}}$ and
$\boldsymbol{\Sigma}^{-1}_{\mathrm{1NN}}$ appearing in
\eqn{eq:chisq_differentiate_defn} are identical to those computed earlier in \secn{subsec:method_detection_prospects} and shown in \fig{fig:corr_matrices}. In this context, $\chi^2_{\mathrm{S};\,(\mathrm{fid},\,\mathcal{M})}$, computed for a given summary statistic $S$, quantifies the degree to which the fiducial model and the chosen model $\mathcal{M}$ differ relative to the expected statistical uncertainties. A high value of $\chi^2_{\mathrm{S};\,(\mathrm{fid},\,\mathcal{M})}$ indicates that the summary statistic $S$ is able to effectively distinguish between the two models. Conversely, a low value of $\chi^2_{\mathrm{S};\,(\mathrm{fid},\,\mathcal{M})}$ indicates that the models are statistically consistent within the expected measurement uncertainties.

\begin{figure}[htbp]
    \centering
    \begin{subfigure}{\columnwidth}
        \centering
        \includegraphics[width=0.98\columnwidth]{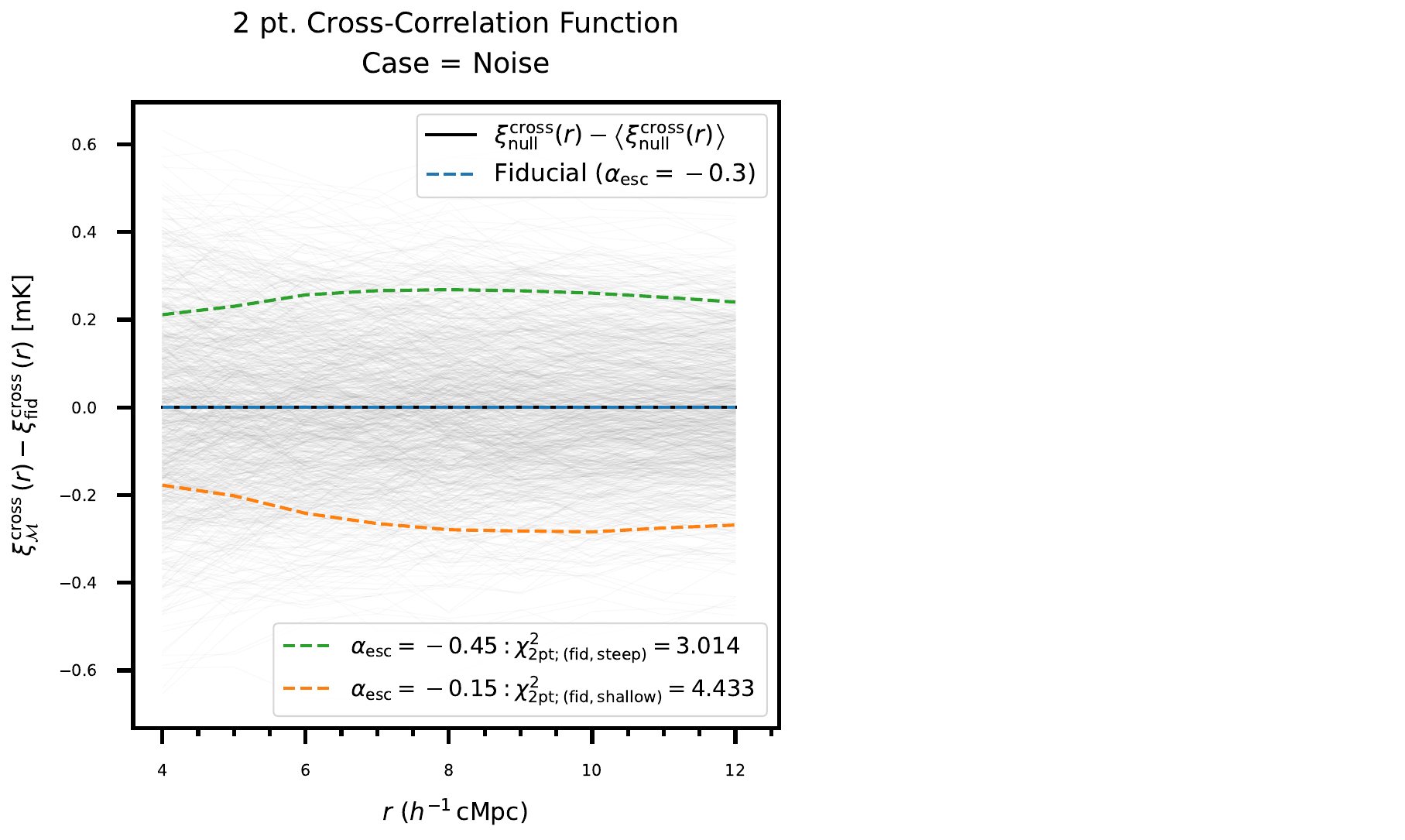}
    \end{subfigure}

    \vspace{0.5cm}

    \begin{subfigure}{\columnwidth}
        \centering
        \includegraphics[width=\columnwidth]{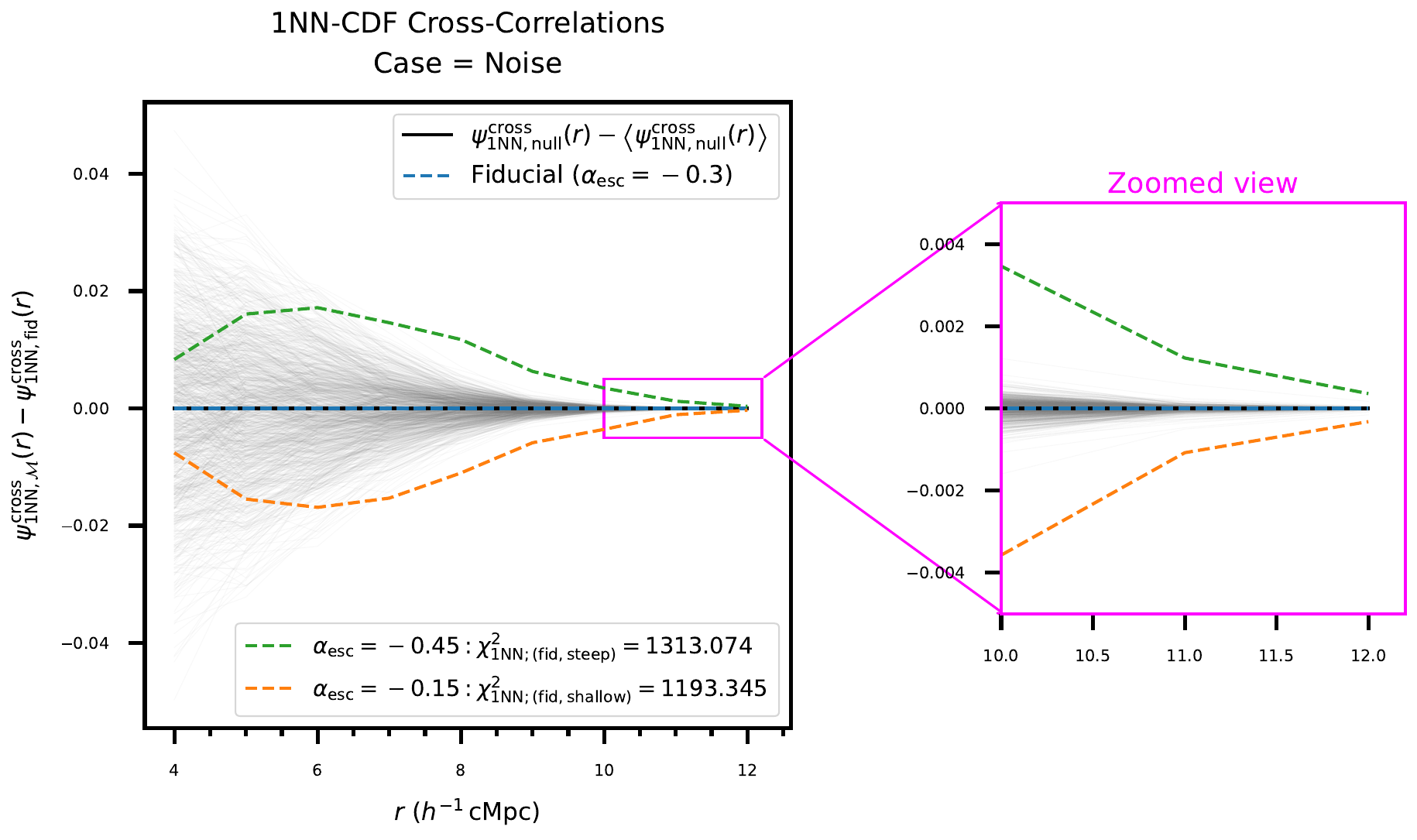}
    \end{subfigure}

    \caption{Comparison of the ability of the two-point cross-correlation function $\xi^{\mathrm{cross}}$ (top panel) and the 1NN CDF cross-correlation statistic $\psi^{\mathrm{cross}}_{1\mathrm{NN}}(r)$ (bottom panel) to distinguish between the contrasting reionization scenarios described in \secn{sec:diff_reion_models} (see \fig{fig:reion_model_comparision}) for the \textbf{Noise} scenario.}
    \label{fig:2pcf_1nn_comparison_noise}
\end{figure}

\begin{figure}[htbp]
    \centering
    \begin{subfigure}{\columnwidth}
        \centering
        \includegraphics[width=0.98\columnwidth]{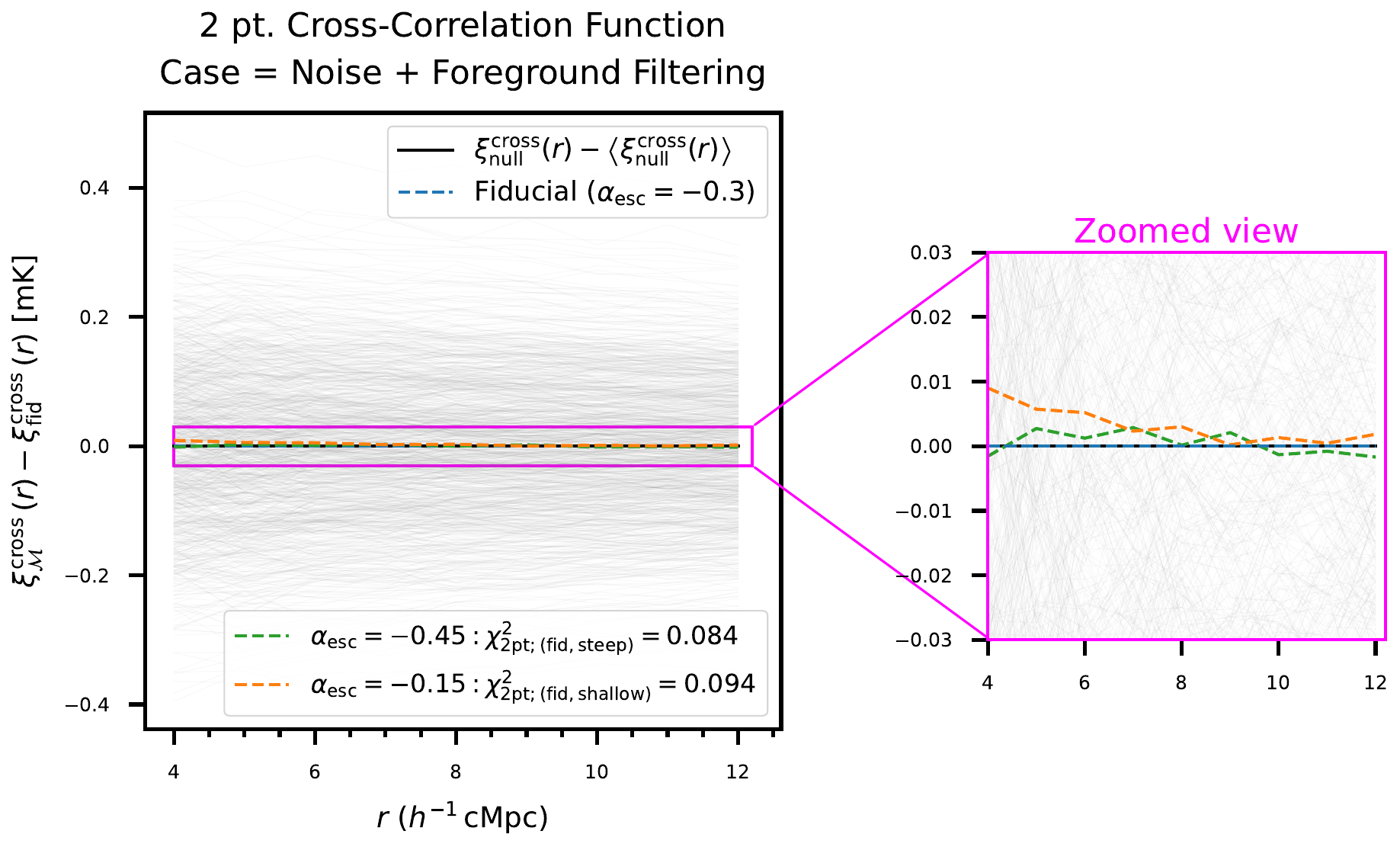}
    \end{subfigure}

    \vspace{0.5cm}

    \begin{subfigure}{\columnwidth}
        \centering
        \includegraphics[width=\columnwidth]{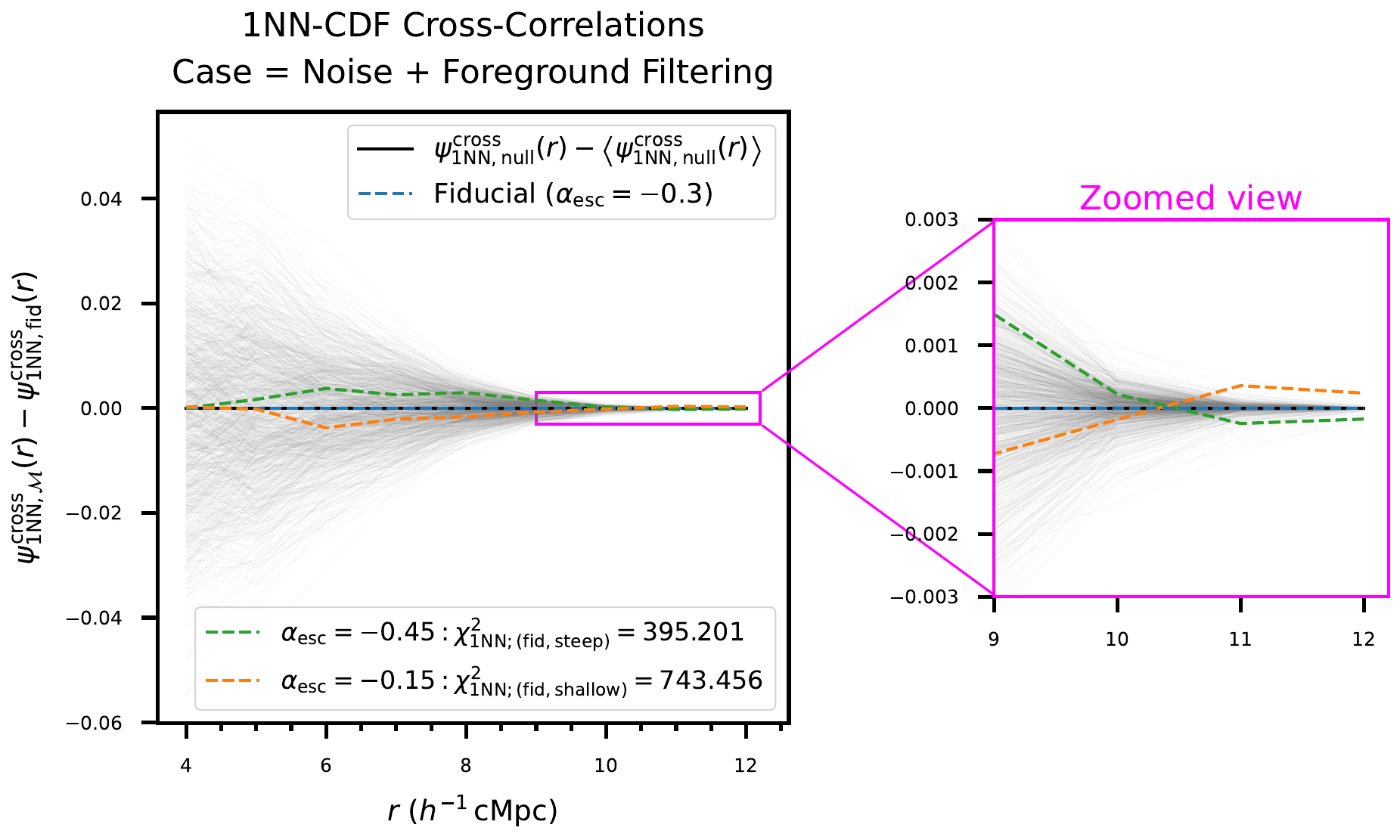}
    \end{subfigure}
    \caption{Same as \fig{fig:2pcf_1nn_comparison_noise}, but for the \textbf{Noise + Foreground Filtering} scenario.}
    \label{fig:2pcf_1nn_comparison_noise_wedge}
\end{figure}

\fig{fig:2pcf_1nn_comparison_noise} illustrates the ability of the two-point 21\,cm--galaxy cross-correlation function and the 1NN CDF cross-correlation statistic to discriminate between the reionization scenarios described above for noise-contaminated 21\,cm fields. \fig{fig:2pcf_1nn_comparison_noise_wedge} extends this comparison to a more realistic observational setting by additionally incorporating the effects of foreground filtering on 21\,cm fields. In both figures, the top panel shows the two-point cross-correlation function, whereas the bottom panel corresponds to the 1NN CDF cross-correlation statistic, evaluated for all three reionization scenarios considered here.

We first discuss the results obtained for the \textbf{Noise} case (\fig{fig:2pcf_1nn_comparison_noise}). The amplitude of the two-point cross-correlation function, $\xi_{\rm cross}(r)$, shows systematic deviations in both the \textbf{steeper-$\bm{f_{\rm esc}}$} and \textbf{shallower-$\bm{f_{\rm esc}}$} models relative to the fiducial scenario, reflecting the sensitivity of two-point statistics to variations in the underlying ionization morphology. However, these model-to-model differences are small compared to the statistical uncertainties of the measurement, as evidenced by the dense ensemble of null realizations. As a result, both alternative models are statistically indistinguishable from the \textbf{fiducial} model, with $\chi^2_{\mathrm{2pt;,(fid,steep)}} \approx 3$ and $\chi^2_{\mathrm{2pt;,(fid,shallow)}} \approx 4$ for the \textbf{steeper-$\bm{f_{\rm esc}}$} and \textbf{shallower-$\bm{f_{\rm esc}}$} cases, respectively.

In contrast, the 1NN CDF cross-correlation statistic, which encodes higher-order morphological information inaccessible to the two-point function, reveals strong deviations from the fiducial model.  Within this framework, the differences between the cross-correlation signals of the alternative reionization scenarios and the fiducial model substantially exceed the statistical scatter over a wide range of intermediate to large scales, as shown in the bottom row of \fig{fig:2pcf_1nn_comparison_noise}. Quantitatively, the \textbf{steeper-$\bm{f_{\rm esc}}$} model is robustly distinguished from the fiducial case with high statistical significance, yielding $\chi^2_{\mathrm{1NN;,(fid,steep)}} \approx 1313$, while the \textbf{shallower-$\bm{f_{\rm esc}}$} model is similarly distinguished with $\chi^2_{\mathrm{1NN;,(fid,shallow)}} \approx 1193$. This demonstrates that the 1NN CDF cross-correlation formalism has considerable discriminatory power with respect to reionization topology, even in the presence of instrumental noise.

We next consider the more realistic \textbf{Noise + Foreground Filtering} case (\fig{fig:2pcf_1nn_comparison_noise_wedge}). As discussed previously, the amplitudes of both cross-correlation statistics are strongly suppressed for all models after foreground filtering -- particularly on large scales -- owing to the removal of long-wavelength modes from the filtered 21\,cm field. For the two-point cross-correlation function, this suppression renders the fiducial and alternate reionization models indistinguishable across the full range of separations, with the residual model-to-model differences falling well within the statistical scatter measured from null realizations. Consistent with this, the chi-square analysis yields very low values of $\chi^2_{\mathrm{2pt;,(fid,steep)}} \approx 0.08$ and $\chi^2_{\mathrm{2pt;,(fid,shallow)}} \approx 0.09$, indicating that the two-point statistic retains little discriminatory power once foreground excision is applied.

The 1NN CDF cross-correlation statistic, on the other hand, remains far more informative even after foreground filtering.
The model-to-model differences in the cross-correlation signal, as measured in this formalism, are clearly detectable at larger separations, as shown in the zoomed panel. In this case, the fiducial model is distinguished from the \textbf{steeper-$\bm{f_{\rm esc}}$} and \textbf{shallower-$\bm{f_{\rm esc}}$} models with a significance quantified by $\chi^2_{\mathrm{1NN;\,(fid,steep)}} \approx 395$ and $\chi^2_{\mathrm{1NN;\,(fid,shallow)}} \approx 743$, respectively.

Overall, we find that while contamination by instrumental noise already limits the power of the two-point cross-correlation function in constraining reionization source models, the inclusion of foreground excision further exacerbates this limitation. In contrast, the 1NN CDF cross-correlation statistic turns out to be consistently more sensitive to variations in the ionization morphology, being able to clearly distinguish between the different reionization scenarios in both observational setups. These results highlight the promise of nearest-neighbour–based statistics as a complementary and potentially more powerful probe of reionization physics in realistic 21\,cm observations affected by foreground contamination.


\section{Conclusion}
\label{sec:conclusion}

The cross-correlation between high-redshift galaxies and the 21\,cm signal from neutral hydrogen in the intergalactic medium promises to be a powerful complementary probe of the Epoch of Reionization (EoR). Such 21\,cm--galaxy synergies are expected not only to help establish the cosmological origin of the observed 21\,cm signal and mitigate residual instrumental and observational systematics, but also to provide independent insights into the growth and topology of ionized regions, as well as the properties of the sources driving reionization

However, as reionization proceeds, the EoR 21\,cm signal becomes increasingly non-Gaussian, rendering a substantial fraction of the underlying physical information inaccessible to traditional two-point cross-correlation statistics. Motivated by these considerations, we present a \emph{proof-of-concept} study investigating the utility of $k$-nearest-neighbour cumulative distribution functions ($k$NN CDFs) as a novel framework for measuring 21\,cm--galaxy cross-correlations during reionization. Using self-consistently simulated mock 21\,cm fields generated with the semi-numerical reionization code \texttt{SCRIPT}, together with a catalog of $\OIII$-emitting galaxies at $z=7$ calibrated to the latest JWST observations, we systematically compare the performance of the $k$NN CDF framework with that of the conventional two-point cross-correlation function under a range of realistic 21\,cm observational setups, including instrumental noise contamination and foreground filtering. In this work, we focus on the $k =1$ case of the $k$NN CDF formalism.

Our main findings can be summarized as follows:
\begin{itemize}
    \item The $1$NN–CDF cross-correlation statistic is able to consistently detect 21\,cm--galaxy cross-correlations at high statistical significance, even when the 21 cm fields are contaminated by thermal noise and/or subject to foreground filtering. In contrast, the two-point cross-correlation function yields only marginal detections in the presence of instrumental noise and becomes statistically consistent with the null hypothesis once foreground filtering is applied. This demonstrates the robustness of the nearest-neighbour cross-correlation formalism to the unavoidable loss of large-scale modes introduced by foreground mitigation strategies.
    
    \item  We further examine the ability of these two statistical frameworks to distinguish between different reionization source models at a fixed global ionized fraction.  We considered a set of models with different halo-mass scalings of the ionizing photon escape fraction, which lead to distinct ionization morphologies, as shown in \fig{fig:reion_model_comparision}. The two-point cross-correlation function fails to discriminate between these models, with the model-to-model differences falling well within the statistical measurement uncertainties. The $1$NN CDF formalism, however, robustly distinguishes between them, even in the presence of foreground filtering, establishing the nearest-neighbour formalism as a powerful tool for both detecting 21\,cm--galaxy cross-correlations as well as constraining the astrophysics of reionization.
\end{itemize}

While our results demonstrate the promise of nearest-neighbour statistics for 21\,cm--galaxy cross-correlation studies during the EoR, fully consolidating their potential will require extending the present proof-of-concept analysis in several important directions. First, our study is restricted to a single redshift ($z=7$) and a specific galaxy tracer population ($\OIII$ emitters). Extending this framework to multiple redshifts and to other tracers that are particularly sensitive to the ionization state of the surrounding medium -- such as Lyman-$\alpha$ emitters -- would enable a more comprehensive assessment of its robustness and its ability to capture the evolving galaxy–IGM connection across different stages of reionization. Second, the current analysis neglects several physical effects that can modify both the galaxy distribution and the 21\,cm signal. We do not model redshift-space distortions arising from peculiar velocities, which can alter the amplitude and scale dependence of the cross-correlation signal by distorting structures along the line of sight. Incorporating these effects will be important for determining how nearest-neighbour statistics perform, relative to other approaches, in detecting the 21\,cm--galaxy cross-correlation signal and in their sensitivity to different reionization scenarios. Similarly, this work assumes coeval volumes and does not include light-cone evolution across the observed redshift interval; accounting for this effect would enable a self-consistent treatment of the evolving ionization morphology and ionization history. In addition, on the galaxy survey side, properly accounting for galaxy selection functions and survey incompleteness will be required to accurately model the observed galaxy distribution used in the cross-correlation analysis. Finally, it will also be useful to compare the $k$NN CDF formalism with other non-Gaussian summary statistics, such as the cross bispectrum, to better understand their relative information content and complementarity for constraining the astrophysics of reionization.

\section*{Acknowledgments}
We thank the organizers and participants of the ``Pune-Mumbai Cosmology and Astroparticle Physics (PMCAP) Meeting'' held at TIFR Mumbai, during which this project was initiated. AC and TRC acknowledge support from the Department of Atomic Energy, Government of India, under project no. 12-R\&D-TFR-5.02-0700. AB's work was partially supported by grants SRG/2023/000378 and ANRF/ARGM/2025/000301/TS from the Anusandhan National Research Foundation (ANRF) India.

\section*{Data Availability}

The data generated and presented in this paper will be made available upon reasonable request to the corresponding author (AC).

\appendix
\section{Incorporating Higher-Order (\boldmath$k>1$) Nearest Neighbour Contributions in the \boldmath$k$NN--CDF Framework}
\label{appendix:addition_of_2NN_for_xcorr_detection}

\begin{figure}[htbp]
    \centering
    
    \begin{subfigure}{\columnwidth}
        \centering
        \includegraphics[scale=0.47]{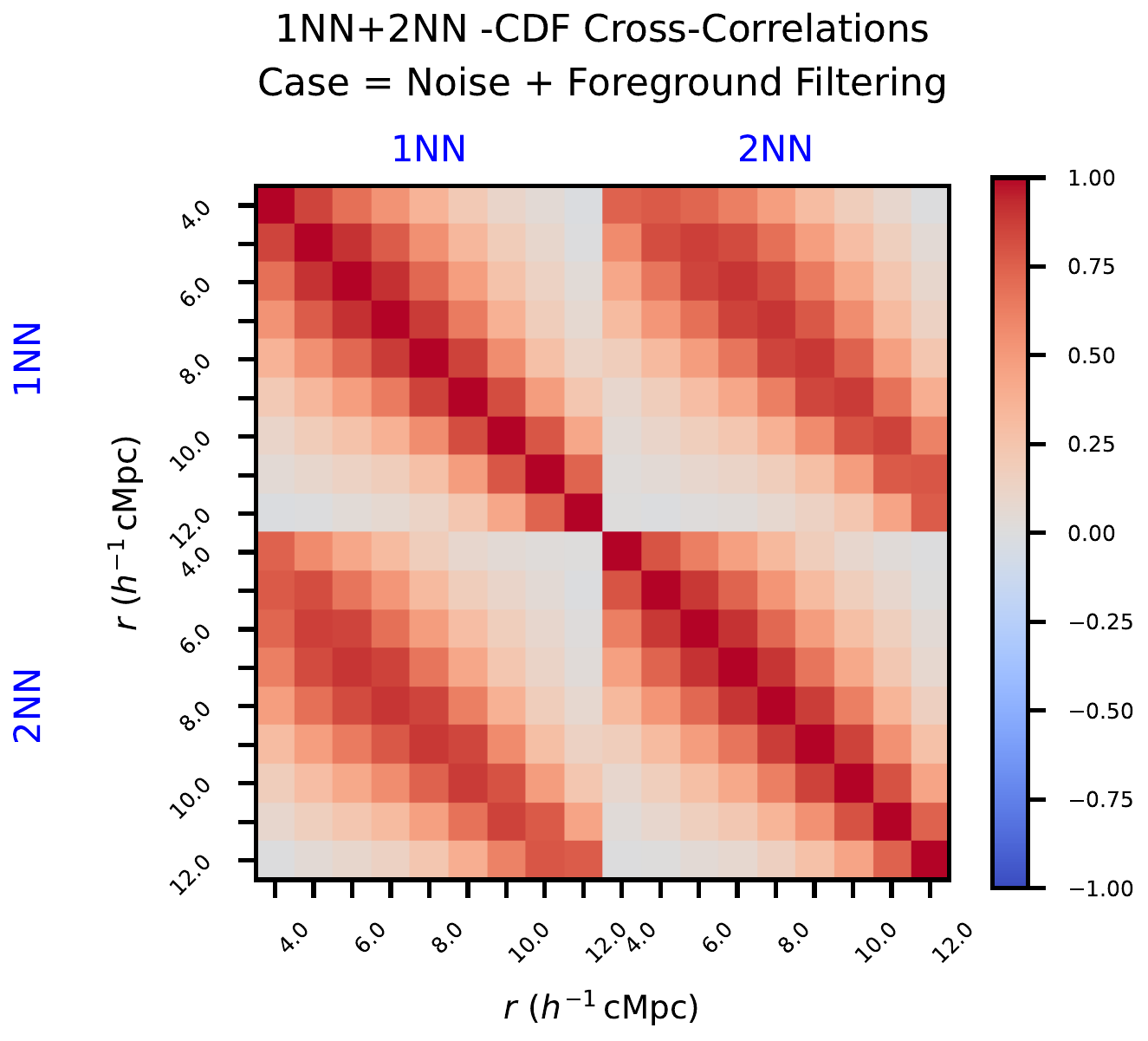}
    \end{subfigure}
    
    \vspace{0.5em}
    
    \begin{subfigure}{\columnwidth}
        \centering
        \includegraphics[scale=0.47]{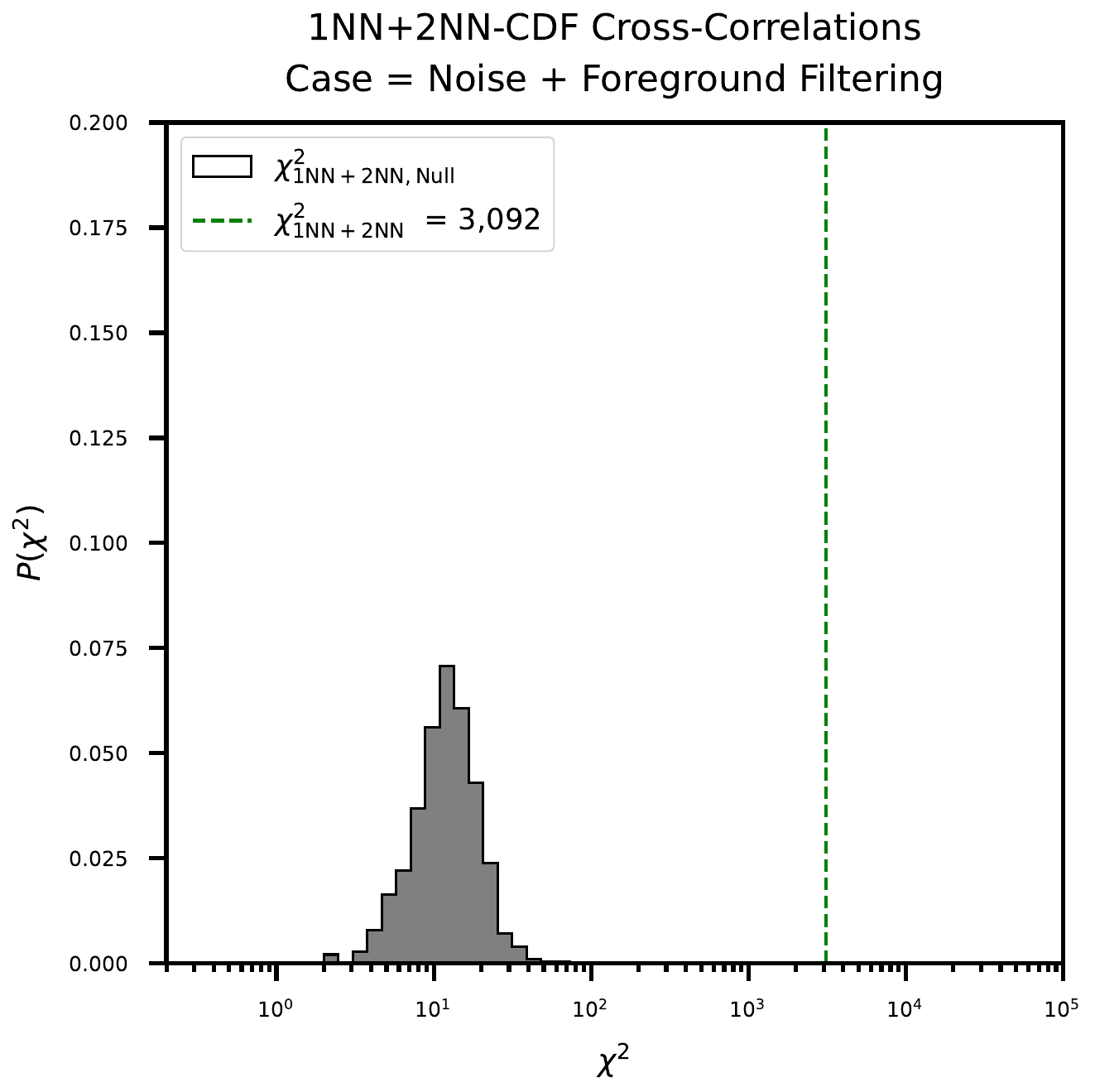}
    \end{subfigure}

    \caption{
    Impact of combining 1NN and 2NN CDFs on the detection of 21\,cm–galaxy cross-correlations in the \textbf{Noise + Foreground Filtering} scenario (see \app{appendix:addition_of_2NN_for_xcorr_detection} for details).
    \textbf{\textit{Top row:}} Correlation matrix of the combined 1NN+2NN cross-correlation statistic, estimated from $N_{\rm null} = 1000$ realizations.
    \textbf{\textit{Bottom row:}} Distribution of $\chi^2$ values obtained for the control samples (gray histograms) and for the mock dataset (green dashed vertical line), using the combined 1NN+2NN cross-correlation statistic.
    }
    \label{fig:detect_prospect_noise_plus_wedge_combinedNN}
\end{figure}

In this appendix, we investigate whether incorporating higher-order nearest-neighbour (NN) information into our cross-correlation analysis improves the prospects for its detection. We focus on the most observationally relevant case, namely the \textbf{Noise + Foreground Filtering} scenario. 

To this end, we construct a joint data vector by combining the first ($k=1$) and second ($k=2$) nearest-neighbour statistics (defined in \eqn{eq:psi_kNN_def}) as follows:
\begin{equation}
\begin{aligned}
\boldsymbol{\psi}^{\rm cross}_{\rm 1NN+2NN}
&\equiv
\left\{
\boldsymbol{\psi}^{\rm cross}_{\rm 1NN}
\, , \,
\boldsymbol{\psi}^{\rm cross}_{\rm 2NN}
\right\} \\[0.5ex]
&=
\left( \,
\psi^{\rm cross}_{\rm 1NN}(r_1), \dots, \psi^{\rm cross}_{\rm 1NN}(r_m), \,
\psi^{\rm cross}_{\rm 2NN}(r_1), \dots, \psi^{\rm cross}_{\rm 2NN}(r_m) \,
\right)
\end{aligned}
\end{equation}

The length (i.e., dimension) of the resulting data vector is therefore now twice that used in the analysis presented in \secn{sec:detection_prospects}. We follow the methodology described in \secn{subsec:method_detection_prospects} to quantify the significance of the cross-correlation signal measured using the combined 1NN+2NN statistic. The results from this analysis are shown in \fig{fig:detect_prospect_noise_plus_wedge_combinedNN}. The top panel displays the structure of the correlation matrix, while the bottom panel shows the distribution of $\chi^2$ values for the control samples, along with the value corresponding to the mock dataset. A comparison of the detection prospects for the different cross-correlation statistics considered in this work is also presented in \tab{tab:stats_summary}.

We find that extending the analysis from the 1NN statistic to the combined 1NN+2NN case increases the total $\chi^2$ from 2028 to 3092. However, this is accompanied by a corresponding increase in the dimensionality of the data vector (from $\nu = 9$ to $\nu = 18$). As a result, the reduced $\chi^2$ decreases from $\approx 225$ to $\approx 171$. This indicates that, although the inclusion of $\boldsymbol{\psi}^{\rm cross}_{\rm 2NN}$ captures additional signal, this gain is counterbalanced by the increased number of degrees of freedom, resulting in no significant improvement in detection prospects.

\begin{table} [htbp]
\centering
\caption{Detection prospects for 21\,cm-galaxy cross-correlations across different statistical frameworks and their corresponding $\chi^2$ values.}
\label{tab:stats_summary}
\begin{tabular}{l|c|c|c}
\hline 
\textbf{Statistical formalism used} & 
\makecell{\textbf{Dimension of the }\\\textbf{data vector (\boldmath$\nu$)}} & 
\makecell{\boldmath$\chi^2$ \textbf{value}} & 
\makecell{\boldmath $\chi_{\rm red}^2 = \ \chi^2/\nu$ \tablefootnote{Since no model parameters are fitted in our analysis, the number of degrees of freedom is equal to the dimension of the data vector.}} \\[0.75ex]
\hline 
2-point statistics : \{$\boldsymbol{\xi}_{\mathrm{cross}}$  \} &  9 &  26   &  2.88  \\[0.75ex]
1NN statistics : \{$\boldsymbol{\psi}^{\rm cross}_{\rm 1NN}$ \} &   9 &  2028 &  225.33 \\[0.75ex]
1NN + 2NN statistics :  \{$\boldsymbol{\psi}^{\rm cross}_{\rm 1NN} \,,\, \boldsymbol{\psi}^{\rm cross}_{\rm 2NN}$ \}&  18 &  3092 &  171.77 \\[0.75ex]
\hline
\end{tabular}
\end{table}


\newpage

\bibliographystyle{JHEP}
\bibliography{manuscript_ver2}

\end{document}